\documentclass[10pt]{iopart}
\usepackage{iopams}
\expandafter\let\csname equation*\endcsname\relax
\expandafter\let\csname endequation*\endcsname\relax
\usepackage{amsmath}
\usepackage{amsfonts}
\usepackage{amsthm}
\usepackage{dsfont}
\usepackage{graphicx}
\usepackage{cite}

\usepackage{bm}

\newcommand{\ud}{\mathrm{d}}
\newcommand{\ui}{\mathrm{i}}
\newcommand{\ue}{\mathrm{e}}

\newcommand{\R}{\mathds{R}}

\newcommand{\cS}{{\mathcal S}}
\newcommand{\cA}{{\mathcal A}}
\newcommand{\cB}{{\mathcal B}}

\providecommand{\norm}[1]{\lVert#1\rVert}

\providecommand{\abs}[1]{\lvert#1\rvert}

\newcommand{\pa}{\partial}
\newcommand{\la}{\langle}
\newcommand{\ra}{\rangle}

\renewcommand{\Im}{\operatorname{Im}}

\newcommand{\m}{D}

\newtheorem{thm}{Theorem}[section]
\newtheorem{lem}[thm]{Lemma}

\newcommand{\phiL}{\phi_{\Lambda}}

\begin{document}

\title{How do wave packets  spread? Time evolution on Ehrenfest time scales. }

\author{Roman Schubert$^1$, Ra\'ul O. Vallejos$^{2}$ and Fabricio Toscano$^3$} 

\address{$^1$School of Mathematics, University of Bristol, 
    University Walk, Bristol BS8 1TW, UK }
\address{$^2$Centro Brasileiro de Pesquisas F\'{\i}sicas (CBPF), 
    Rua Dr.~Xavier Sigaud 150, 
    22290-180 Rio de Janeiro, Brazil}
\address{$^3$Instituto de F\'{\i}sica,
    Universidade Federal do Rio de Janeiro,
    Cx.~P. 68528, 21941-972 Rio de Janeiro, Brazil}

\begin{abstract}
We derive an extension of the standard time dependent WKB theory which can be applied to 
propagate coherent states and other strongly localised states for long times. It allows in particular 
 to give a uniform description of  the transformation  from 
a localised coherent state to a delocalised Lagrangian state which takes place at the Ehrenfest time. 
The main new ingredient is a metaplectic operator which is used to modify the initial state in a way that 
standard time dependent WKB can then be applied for the propagation. 

We give a detailed analysis of the phase space geometry underlying this construction and  use 
this to determine the range of validity of the new method. Several examples are used to illustrate and test 
the scheme and two applications are discussed: (i) For scattering of a wave packet on a barrier near the critical energy we can derive 
uniform approximations for the transition from reflection to transmission.   (ii) 
A wave packet propagated along a hyperbolic trajectory becomes a Lagrangian state 
associated with the unstable manifold at the Ehrenfest time, this is illustrated with the kicked harmonic oscillator.

\end{abstract}

\pacs{03.65.Sq, 05.45.Mt, 82.20.Ln}



\section{Introduction}

The semiclassical propagation of wave packets is of fundamental importance 
in  many applications of quantum mechanics and has therefore 
been studied extensively in the literature \cite{BerMou72,Ber79,Lit86,Hel91,Lit92,YeaUze00,Ali00,Tan07,TosValWis09}. Time dependent WKB 
approximations apply to extended initial states of the form 
\begin{equation}\label{eq:WKBAns}
\psi(x)=A(x)\ue^{\frac{\ui}{\hbar}S(x)}
\end{equation}
where the phase function $S(x)$ is real valued and smooth and 
the amplitude $A(x)$ is smooth and non-oscillating. The time evolution of 
such states can  be expressed in terms of transport along 
a family of classical trajectories determined by initial positions $x$ and 
initial momenta $p=\nabla S(x)$, see, e.g., \cite{Ber79,Lit92}. 

A different class of initial states is given by coherent states, see, e.g.,  \cite{Lit86,Ali00},  which are strongly localised 
and therefore their propagation can be described up to certain times using 
only one trajectory and the linearised motion around it. 
A Gaussian  coherent state is a function of the form 
\begin{equation}\label{eq:coh_state_def}
\psi_Z^B(x)=\frac{(\det \Im B)^{1/4}}{(\pi\hbar)^{n/4}}\ue^{\frac{\ui}{\hbar}[p\cdot (x-q)+(x-q)\cdot B
(x-q)/2]}
\end{equation}
where $Z=(p,q)\in \R^{2n}$ is a set of parameters which determine where 
the state is localised in phase space and $B$ is a complex symmetric $n\times n$ matrix with 
$\Im B>0$. The latter condition ensures that the state is localised around $x=q$. 
The roles of $B$ and $Z$ become clearer if we look at the Wigner function of 
such a state, which is  a Gaussian of the form 
\begin{equation}
W(z)=\frac{1}{(\pi\hbar)^n}\ue^{-(z-Z)\cdot G (z-Z)/\hbar}\,\, ,
\end{equation}
with phase space coordinates $z\in\R^n\times \R^n$ and where  $G$ is a positive definite symmetric and symplectic matrix
determined by $B$. 
As was described by Hepp and Heller \cite{Hep74,Hel75,Hel76,DavHel84}, the time evolution of 
a coherent state can in the semiclassical limit be described by using just 
the classical trajectory $Z(t)=(p(t),q(t))$ through $Z$ and  
the  linearised flow around it which is a time dependent 
$2n\times 2n$ symplectic matrix $\cS(t)$. 
This simple description of the semiclassical propagation of coherent states made them a very useful tool in many 
applications, in particular in chemistry due to the work of Heller and coworkers and their appearance in initial value 
representations like the Herman-Kluk propagator \cite{Lit86,Hel91,DavHel84,Hag80,Hel81, HerKlu84,  HubHelLit88,Kay94}

We will be using time dependent WKB methods developed for states of the form \eqref{eq:WKBAns} to understand the propagation of the states \eqref{eq:coh_state_def} for long times.  
In order to identify 
the relevant time scales it is useful to consider 
 the Wigner function of the time evolved coherent state which in  leading order is given  by 
\begin{equation}\label{eq:propWig}
W(t,z)\approx \frac{1}{(\pi\hbar)^n}\ue^{-(z-Z(t))\cdot G(t)(z-Z(t))/\hbar}
\end{equation} 
where $G(t)= \cS^{\dagger}(t)G\cS(t)$ and $\cS(t)$ is  the linearised flow around the central trajectory $Z(t)$. 
The basic idea leading to this approximation  is that since a coherent state is 
localised around a point in phase space one can approximate 
the Hamiltonian by its  quadratic Taylor expansion  around the 
classical trajectory $Z(t)$. The resulting  Schr{\"o}dinger 
equation can then be solved explicitly. 
This approximation can only be expected to be accurate as long as the 
propagated state stays localised, and from  \eqref{eq:propWig} we see that 
this is the case  as long as $\lambda_{min}[G(t)]/\hbar\gg 1$, where we denote by 
$\lambda_{min/max}[G(t)]$  the smallest or largest  eigenvalue of the matrix $G(t)=\cS^{t}(t)G\cS(t)$, respectively. 
Since $G(t)$ is symmetric and positive we have  $ \lambda_{min}[G(t)]>0$, and
that $G(t)$ is symplectic implies  that
$1/ \lambda_{min}[G(t)]=\lambda_{max}[G(t)]$, so the localisation condition becomes 
\begin{equation}
\lambda_{max}[G(t)]\hbar\ll 1\,\, .
\end{equation}
In order to relate this more explicitly to the properties of the classical dynamics we use the estimate 
$\lambda_{max}[G(t)]=\norm{G(t)}\leq C \norm{\cS(t)}^2$ and so the condition becomes  $\norm{\cS(t)}\sqrt{\hbar}\ll 1$. Here 
we use for a matrix $A$ the norm induced by the 
Euclidean norm $\norm{x}=\sqrt{ x\cdot x}$, i.e.,  $\norm{A}:=\sup_{\norm{x}=1} \frac{\norm{Ax}}{\norm{x}}$. 
The time at which the width of a propagated coherent  state reaches order 
one is called the Ehrenfest time $T_E(\hbar)$, and by the previous discussion we see that it is determined by the condition 
\begin{equation}
\norm{\cS(T_E(\hbar))}\sqrt{\hbar}= 1\,\, .
\end{equation}
Hence the Ehrenfest time depends on the dynamical properties of the classical system  near the trajectory $Z(t)$. 
 If the trajectory $Z(t)$ is hyperbolic, with largest Liapunov exponent 
$\lambda>0$, then $\norm{S(t)}\sim \ue^{\lambda t} $ and so 
\begin{equation}
T_E(\hbar)=\frac{1}{2\lambda}\ln{\frac{1}{\hbar}}\,\, , 
\end{equation}
on the other hand side, if the dynamics is integrable  we   typically have $\norm{\cS(t)}\sim t$ for large $t$ and then  
\begin{equation}\label{eq:Ehrenfest_integrable}
T_E(\hbar)=\frac{1}{\sqrt{\hbar}}\,\, . 
\end{equation}
Ehrenfest showed in his classical paper \cite{Ehr27} how one recovers classical mechanics from quantum dynamics
 by considering expectation values in propagated localised states. His 
construction breaks down if the state becomes delocalised, hence the name Ehrenfest time for 
the time scale at which this happens. 
In more  recent years the behaviour of  time evolved coherent states for long times has
attracted the attention of mathematicians and rigorous estimates for long times have been derived,  
in particular in \cite{ComRob97,HagJoy00} it was shown that 
the remainder terms stays small for times satisfying  $\abs{t}< \frac{1}{3}T_E$ in the hyperbolic case. 
One should emphasise however that the breakdown of Ehrenfest's  argument at the Ehrenfest time does not mean 
that the quantum to  classical correspondence breaks down, it only means that the approximate propagation of 
a wave packet based on one central trajectory breaks down. If one includes more trajectories one can use 
semiclassical propagation well beyond the Ehrenfest time as has been demonstrated in \cite{TomHel91,TomHel93}, 
although from a mathematical point of view this is still a challenging problem.

The principal aim of this work is to present a propagation scheme for coherent states which works at    
times of the order of  the Ehrenfest time $t\sim T_E$, and is able to describe the transition from 
a localised state \eqref{eq:coh_state_def} to an extended state \eqref{eq:WKBAns} at the Ehrenfest time in a uniform way. 
 In a previous paper \cite{MaiNicValTos08} it was  demonstrated that at the Ehrenfest time a coherent state which is
transported along a hyperbolic trajectory  
becomes effectively a WKB state of the type \eqref{eq:WKBAns}  associated with the unstable manifold 
of that trajectory. Here we will give the theoretical
foundation for that claim presenting a new propagation scheme which 
is a combination of a time dependent WKB approximation and a metaplectic correction. 

The qualitative change in the nature of the state at the Ehrenfest time  
was  described in \cite{BonDeB00} for the cat map and 
in \cite{DeBRob03} for   the propagation of coherent states centred on 
unstable fixed points in one-dimensional 
multiple well potentials. In contrast to the present approach that 
work was not based on direct propagation of the state in the 
position representation but on the Wigner function and the 
Egorov theorem from \cite{BouRob02}. That means in particular that we can give a much more
detailed description of the state at and beyond the Ehrenfest time.

The plan of the paper is as follows. In Section \ref{sec:realTDWKB} we recall time dependent WKB theory 
for the propagation of states of the form \eqref{eq:WKBAns} and rewrite it in an exact  form where we 
explicitly separate the classical transport and the quantum dispersion. The main idea of this paper is then 
developed in  Section \ref{sec:met_ext} where we show how to approximate the 
action of the dispersive part on  a coherent state using a simple metaplectic operator and  combine this with time dependent WKB 
approximation to obtain a method which allows to describe the  propagation of  coherent states at and beyond the Ehrenfest time. 
In Section \ref{sec:geom_int} we develop the phase space geometry underlying the metaplectically extended WKB 
method. This allows us to determine its range of validity, in particular its dependence on dynamical properties of the classical system. 
We then develop the geometric picture even further in Section \ref{sec:gen_scheme} to allow more general Hamiltonians and 
the inclusion of caustics after the Ehrenfest time, the price we have to pay is that remainder estimates become less explicit now. 
The last two sections are devoted to examples. In Section \ref{sec:exI} we consider a couple of simple explicitly or almost 
explicitly solvable examples which demonstrate in some detail how our method works in practice. We show in particular that 
for a potential barrier we can  describe the transition from transmission to reflection near the critical energy in a uniform way. 
In Section \ref{sec:exII}  we reconsider the situation from \cite{MaiNicValTos08} and demonstrate how our method 
reproduce the results. As a particular application this shows using \cite{Schub05} that in a strongly chaotic system coherent 
states become semiclassically equidistributed beyond the Ehrenfest time. In the end we summarise our results in the conclusions. 
In the appendix we collect some more technical results from semiclassical analysis we use in the main body of the work.

\section{Time dependent WKB}\label{sec:realTDWKB}

Our method is an extension of the well known  time dependent WKB method for real valued phase functions. So we start by recalling 
this method, and  to keep the discussion simple we restrict ourselves first to the 
case of the standard Schr{\"o}dinger equation with potential $V$ in $\R^n$, which in appropriate units reads 
\begin{equation}
\ui\hbar\pa_t \psi=-\frac{\hbar^2}{2}\Delta \psi+V\psi\,\, , 
\end{equation}
in Section \ref{sec:gen_scheme} we will discuss a more general setting. 
 We want to solve this equation for an initial state of the form $\psi=A_0\ue^{\frac{\ui}{\hbar}S_0}$, where $S_0$ is real valued. 
Inserting the usual WKB ansatz 
\begin{equation}\label{eq:tdepWKB}
\psi(t,x)=A(t,x)\ue^{\frac{\ui}{\hbar}S(t,x)}\,\, ,
\end{equation}
with  $S(t,x)$ real valued, into the Schr{\"o}dinger equation gives an expression which we 
 separate, following the standard treatment \cite{Lit92},  into  the following two equations 
\begin{align}
\pa_tS+\frac{1}{2}\abs{\nabla S}^2+V&=0\label{eq:HamJac}\\
\ui\pa_tA+\ui\nabla S \cdot\nabla A +\frac{1}{2}\ui (\Delta S) A&=
\frac{\hbar}{2}\Delta A\label{eq:trans_hbar}\,\, .
\end{align}
If $A$ does not depend on $\hbar$, then this is a splitting into different
powers of $\hbar$. 
 Equation \eqref{eq:HamJac}  is the Hamilton-Jacobi
equation and its solutions are described using the propagation of 
the Lagrangian manifolds defined by $p=\nabla S(t,q)$, i.e., 
\begin{equation}
\Lambda_t:=\{(\nabla S(t,x),x)\}\subset \R^n\times \R^n\,\, ,
\end{equation}
which are transported by the Hamiltonian flow associated with the classical 
Hamiltonian $H(p,q)=\frac{1}{2}p^2+V(q)$. We will analyze this further in Section \ref{sec:geom_int}. 
The phase function $S(t,x)$ often exists only 
on a finite time interval and develops singularities if caustics appear, then one has to 
use a refined Ansatz.  Because of the 
association with Lagrangian submanifolds WKB states of the form \eqref{eq:tdepWKB}
are often called Lagrangian states in the mathematical literature \cite{DimSjo99,Dui11}. 

If we ignore in \eqref{eq:trans_hbar} the term of order $\hbar $ on the right
hand side, then this 
is a transport equation which describes  how the amplitude $A$ is transported along the
Lagrangian manifold $\Lambda_t$. Let us define the transport operator $T(t)$ as 
the solution to the transport equation  
\begin{equation}\label{eq:defT}
\ui \pa_t T(t)=-\bigg[\ui\nabla S\cdot \nabla +\ui\frac{1}{2} \Delta S\bigg]T(t)\,\, ,
\quad T(0)=I\,\, ,
\end{equation}
where the operator in brackets on the right hand side is self-adjoint, hence $T(t)$ is unitary. 
Then if we make for $A(t)$ an ansatz
\begin{equation}
A(t)=T(t)D(t)A_0\,\, ,
\end{equation}
with an operator $D(t)$ which is to be determined, and insert this into \eqref{eq:trans_hbar}, then 
we obtain  for $D(t)$ the equation
\begin{equation}\label{eq:D_eq}
\ui\pa_t D(t)=-\frac{\hbar}{2}\Delta(t)D(t)\,\, ,\end{equation}
where 
\begin{equation}\label{eq:D_gen}
\Delta(t):=T^*(t)\Delta T(t)\,\, ,
\end{equation}
and with initial condition $D(0)=I$. Since $\Delta(t)$ is self-adjoint, $D(t)$ is unitary as well. From a more detailed analysis of the operator $T(t)$ in Section \ref{subsec:transport} we will learn that  $\Delta(t)$ is a generalised Laplacian of the form 
\begin{equation}\label{eq:delta_t_expl}
\Delta(t)=\sum \alpha_{ij}(t,x)\pa_i\pa_j+\sum \beta_i(t,x)\pa_i\,\, ,
\end{equation}
see \eqref{eq:symb_deltat}.
Collecting all terms  we have a solution to the original Schr{\"o}dinger equation 
of the form 
\begin{equation}\label{eq:TDS}
\psi(t)=[T(t)D(t)A_0]\ue^{\frac{\ui}{\hbar}S(t)}\,\, .
\end{equation}
So the time evolution is described by three parts, (i) propagation 
of the Lagrangian manifold $\Lambda_t$, (ii) classical transport of the amplitude 
by the unitary operator $T(t)$ and (iii) quantum dispersion described by 
$D(t)$. Note that both $T(t)$ and $D(t)$ depend on the initial state via the initial phase function 
$S_0$.

The main condition we need in order that  \eqref{eq:TDS} holds on a time interval $[0,T]$ is that the projection of $\Lambda_t$ to position space is smooth  for all $t\in [0,T]$, i.e., that $\Lambda_t$ does not develop caustics. 
Under this condition  the representation   \eqref{eq:TDS} is exact, and $S(t)$ and $T(t)$ are both determined by 
transport along classical trajectories in phase space, the only contribution 
not yet linked to classical dynamics is the dispersive part $D(t)$. 
We get the usual WKB result by neglecting the dispersion, i.e., 
by replacing $D(t)$ by the identity $I$.  To  see 
what kind of error we make by doing this we can use Duhamel's principle which we will 
state in  general form for later use as well \cite{ComRob97}. Let $\hat H(t)$ and $\hat H_1(t)$ be two time dependent 
self-adjoint operators, and $U(t)$ and $U_1(t)$ the time evolution operators
generated by 
them with initial conditions 
$U(0)=U_1(0)=I$, then
\begin{equation}\label{eq:Duhamel}
U(t)-U_1(t)=-\frac{\ui}{\hbar}\int_0^t U(t)U^*(s)[\hat H(s)-\hat H_1(s)]U_1(s)\, \ud s\,\, .
\end{equation}
This formula allows to estimate how close the time evolutions generated by two different Hamiltonians 
are to each other. If we choose  
$\hat H(t)=-\frac{\hbar^2}{2}\Delta(t)$ 
and $\hat H_1=0$  then 
$U(t)=D(t)$ and $U_1(t)=I$, and  \eqref{eq:Duhamel} gives 
\begin{equation}
D(t)A_0=A_0+\frac{\ui\hbar}{2}\int_0^t D(t)D^*(t')\Delta(t')A_0 \, \ud t'\,\, .
\end{equation}
Since $D$ is unitary we obtain then directly 
\begin{equation}\label{eq:D-I_est}
\norm{D(t)A_0-A_0}\leq \frac{\hbar}{2}\int_0^t\norm{\Delta(t')A_0}\, \ud t'\,\, .
\end{equation}
So if the second order derivatives of $A_0$ are bounded and the 
coefficients of $\Delta(t)$ are not growing with $t$, then the solution to
Schr{\"o}dinger's equation satisfies. 
\begin{equation}
\psi(t)=(T(t)A_0)\ue^{\frac{\ui}{\hbar}S(t)}+O(\abs{t}\hbar)
\end{equation}
 which is the standard time dependent WKB result. 
Furthermore Duhamel's formula can be iterated and gives an expansion 
with error term of order $O((\hbar\abs{ t})^N)$ if the the first 
$2N$ derivatives of $A_0$ are bounded and the coefficients of 
$\Delta(t)$ don't grow with $t$,  \cite{ComRob97}.

\section{Metaplectic extension of WKB}\label{sec:met_ext}

We would like to apply  \eqref{eq:TDS} to a coherent state, i.e., a state of the form \eqref{eq:coh_state_def}. 
With $S_0(x)=p\cdot(x-q)$ and $A_0(x)=(\pi\hbar)^{-n/4}\ue^{-\frac{1}{2\hbar}(x-q)\cdot B(x-q)}$ 
this state is of the form we can use in \eqref{eq:TDS}, but now 
\begin{equation}
\norm{\Delta(t)A_0}\sim 1/\hbar\,\, ,
\end{equation}
so \eqref{eq:D-I_est} does not allow us to use the standard
WKB approximation $D(t)\approx I$.

The way to solve this problem is to approximate the action of $D(t)$ on 
$A_0$ not by the identity, as in the  WKB method, 
but to borrow from the standard propagation of coherent states and 
approximate the generator of $D(t)$ by its Taylor expansion around the centre of $A_0$ up to second
order.  Since $D(t)$ acts on a state which is concentrated in position space  at 
$x=q$ and in momentum space at $p=0$ this means that we  freeze 
the coefficients of $\Delta(t)$ at $x=q$ and approximate $D(t)$ by the 
operator generated by it.

To formalise this idea  we introduce for $q\in\R^n$ 
 the unitary operator $L_q$ by 
\begin{equation}\label{eq:scalingOp}
(L_qa)(x):=\hbar^{-n/4}a((x-q)/\sqrt{\hbar})\,\, ,
\end{equation}
which shifts by $q$ and then rescales in $\hbar$, so that 
$(L_qa)(x)$ is concentrated around $x=q$, e.g., 
if $a(x)=\pi^{-n/4}\ue^{-xBx/2}$ then $A_0=L_qa$ is the amplitude of 
the coherent state \eqref{eq:coh_state_def} considered above. Now inserting $A=L_qa$ into \eqref{eq:D_eq} gives 
for $a$ the equation
\begin{equation}
\ui\pa_t a=\hbar L_q^*\Delta(t)L_q a\,\, ,
\end{equation}
and we will approximate the operator on the right hand side by 
\begin{equation}\label{def:D0}
\Delta_q(t):=\lim_{\hbar\to 0}\hbar L_q^*\Delta(t)L_q  \,\, .
\end{equation}
By \eqref{eq:delta_t_expl}   one  finds that 
 $\Delta_q(t)=\sum \alpha_{ij}(t,q)\pa_i\pa_j$ 
and 
\begin{equation}\label{eq:diffDelta}
\hat Q_q(t):=\hbar L_q^*\Delta(t)L_q-\Delta_q(t)=O_t(\sqrt{\hbar})
\end{equation}
 in the sense that $\hat Q_q(t)a(x)=O_t(\sqrt{\hbar})$ if $a$ is smooth and has bounded derivatives.
So the  operator $\Delta_q(t)$ is indeed obtained from $\Delta(t)$ by freezing the coefficients at $x=q$ and furthermore discarding the first order terms. In \eqref{eq:delta_q_expl} we will obtain  more explicit bounds on \eqref{eq:diffDelta} in terms of the classical flow. 

Therefore if we define the operator $\m_q(t)$ by 
\begin{equation}\label{eq:def_meta}
\ui\pa_t \m_q(t)=-\frac{1}{2}\Delta_q(t)\m_q(t)\,\, ,\quad \m_q(0)=I\,\, ,
\end{equation}
then we expect that $D(t)L_qa\approx L_q\m_q(t)a$ holds.
To check this we use again  Duhamel's principle \eqref{eq:Duhamel}
\begin{equation}
D(t)L_qa= L_q\m_q(t)a+\frac{\ui}{2} 
\int_0^t D(t)D^*(s)L_q\hat Q_q(s) \m_q(s) a\, \ud s
\end{equation}
 and, since $D(t)$ and $L_q$ are unitary, we get from \eqref{eq:diffDelta}
\begin{equation}\label{eq:error_meta}
\norm{D(t)L_qa-L_q\m_q(t)a}\leq\int_0^t  \norm{\hat Q_q(s)\m_q(s)a}\, \ud s =O_t(\sqrt{\hbar})\,\, .
\end{equation}
 The $t$-dependence in the remainder term will be  governed by the 
 behaviour of $\hat Q_q(t)$. In the next section we give conditions on the initial state and on the classical dynamics 
 under which the coefficients of $\hat Q_q(t)$ stay bounded, see \eqref{eq:A_bounded} and \eqref{eq:delta_q_expl}.

To summarise our results so far, if $\psi_0=A_0\ue^{\frac{\ui}{\hbar}S_0}$ where 
$A_0=L_qa$ is concentrated around $q$, then the time evolved state is given by 
\begin{equation}\label{eq:TD_approx}
\psi(t)=[T(t)L_q\m_q(t)a]\ue^{\frac{\ui}{\hbar}S(t)}+O_t(\sqrt{\hbar})\,\, , 
\end{equation}
so we can use standard time dependent WKB to propagate coherent states 
if we include the additional operator $\m_q(t)$. This operator has a quadratic generator and is therefore a 
metaplectic operator, see \cite{Lit86,Fol89,ComRob06}, and so we call it the \emph{metaplectic correction} to the standard time dependent WKB method.
 In order to understand the behaviour of $\m_q(t)$ in some more detail we have to 
look at the propagation of the Lagrangian manifold $\Lambda$ 
by the classical dynamics, and the associated geometric interpretation of 
our extended time dependent WKB scheme. This will be the subject of 
the next section. 

We finally note that it is sometimes useful to work with $M_q(t):=L_q\m_q(t)L^*_q$, 
because then $L_q\m_q(t)a= M_q(t)L_qa= M_q(t) A_0$ and we can allow for 
$A_0$ to be of a more general form than the one induced by the specific scaling with 
$L_q$, as long as it is concentrated around $x=q$. Notice that $ M_q$ satisfies the 
equation
\begin{equation}\label{eq:def_meta_tilde}
\ui\hbar\pa_t M_q(t)=-\frac{\hbar^2}{2}\Delta_q(t) M_q(t)\,\,  ,\quad  M_q(0)=I\,\, ,
\end{equation}
and the time evolution of an initial state  $\psi_0=A_0\ue^{\frac{\ui}{\hbar}S_0}$, 
where $A_0(x)$ is concentrated around $x=q$, can be approximated by 
\begin{equation}\label{eq:TM_approx}
\psi(t)= [T(t) M_q(t)A_0]\ue^{\frac{\ui}{\hbar}S(t)}+O_t(\sqrt{\hbar})\,\, .
\end{equation}

\section{Geometric interpretation}\label{sec:geom_int}

In order to understand the  scope and the accuracy of the metaplectic
extension of time dependent WKB theory we outlined in the previous section, we
have to understand some of the geometry underlying it. We will see that 
in particular the time dependence of the remainder term is 
governed by the way in which the classical flow transports the 
initial Lagrangian manifold $\Lambda$. Furthermore, a proper understanding of
the geometry will allow to extend the scheme beyond  caustics.   

\subsection{Transport of the Lagrangian manifold}

Let us first return to the solutions of the Hamilton Jacobi equation 
\eqref{eq:HamJac}, which we will discuss for a  general Hamiltonian $H(\xi,x)$,  
\begin{equation}
\pa_t S(t,x)+H(\nabla S(t,x),x)=0\,\, .
\end{equation}
As is well known, there are two main 
ingredients involved in the solution, see e.g., \cite{Lit92, DimSjo99, Dui11}.  The first is the transport of the initial Lagrangian 
manifold 
\begin{equation}
\Lambda:=\{(\nabla S(x),x)\, :\,  x\in U\}\,\, ,
\end{equation}
where $U\subset \R^n$ is an open set which contains the support of the amplitude $A(x)$. Let us denote  
by $\Phi^t(\xi,x)$, where $(\xi,x)\in\R^n\times\R^n$, the Hamiltonian flow
generated by $H$, 
i.e., $\Phi^t(\xi,x)=(p(t),q(t))$ where $(p(t),q(t))$ are the solutions to Hamiltons equation 
$\dot p=-\nabla_qH(p,q)$ and $\dot q=\nabla_p H(p,q)$ with initial conditions
$p(t=0)=\xi$ and $q(t=0)=x$, respectively. Using this flow we can transport the Lagrangian manifold 
$\Lambda$ and get a family of Lagrangian manifolds 
\begin{equation}
\Lambda_t=\Phi^t(\Lambda)=\{\Phi^t(\nabla S(x),x)\, :\, x\in U\}\,\, .
\end{equation}

We will call the initial Lagrangian manifold $\Lambda$
\emph{non-contracting} (with respect to the flow $\Phi^t$) if there exists a $C>0$ such that 
for all $t>0$
\begin{equation}\label{eq:def_non_contracting}
\norm{\ud \Phi^t|_{\Lambda}(z)}\geq C \,\, ,\quad\text{for all}\quad z\in\Lambda\,\, .
\end{equation}
Here $\ud \Phi^t|_{\Lambda}(z):T_z\Lambda\to T_{\Phi^t(z)}\Lambda_t$ is the restriction of the 
linearised flow at $z=(p,q)$ to $\Lambda$ which is given in local coordinates by
the  matrix of the  derivatives of the components 
of $\Phi^t$ with respect to the coordinates on $\Lambda$.
This condition means that,  roughly speaking,  
trajectories starting nearby on $\Lambda$ do not coalesce into each other. 
An example for a non-contracting submanifold is the unstable manifold of a
hyperbolic trajectory, 
and more generally, 
if a system is hyperbolic then any submanifold which is transversal to 
the stable foliation is non-contracting. Stable manifolds are then of course examples of manifolds 
which are contracting, and so  not non-contracting. In an integrable system any manifold which lies in the regular 
part of the foliation of phase space into invariant tori is non-contracting.

The second ingredient needed to determine $S(t,x)$ is the projection of $\Lambda_t$ to  position space 
along the momentum directions, i.e., we take the projection $\pi :\R_p^n\times \R_q^n\to \R_q^n$ defined 
by $\pi(p,q)=q$ and restrict it to $\Lambda_t$. If this map, 
\begin{equation}
\pi_{\Lambda_t}:\Lambda_t\to \R^n\,\, , 
\end{equation}
has no singularities, i.e., $\pi(\Lambda_t)=U_t$ does not contain any caustics of
$\Lambda_t$,  then there exist, 
at least locally, a phase function $S(t,x)$ such that 
\begin{equation}\label{eq:genLt}
\Lambda_t=\{(\nabla S(t,q), q)\, :\, q\in U_t\}\,\, ,
\end{equation}
 for some open set $U_t\in\R^n$.
By assumption $\pi_{\Lambda_0}:\Lambda_0\to U_0\subset \R^n$ is smooth so we
can invert it, with inverse  given by $\pi_{\Lambda_0}^{-1}(x)=(\nabla
S(0,x),x)$, 
and hence we can form
\begin{equation}
\phiL (t,x):=\pi_{\Lambda_t}\Phi^t\big(\pi_{\Lambda_0}^{-1}(x)\big)
\end{equation}
which is a map from $U\to U_t$, and if $\pi_{\Lambda_t}$ is smooth, it is a smooth and invertible map. 
But notice that $\phiL(t)$ is not a flow. 
Analogously to the above notion of non-contracting, \eqref{eq:def_non_contracting},   we say that $\phiL(t)$
is \emph{non-contracting} on $U\subset \R^n$ if 
there exist a $C>0$ such that
\begin{equation}\label{eq:F_non_contr}
\norm{ \phiL'(t,x)}\geq C\,\, ,\quad \text{for all}\quad x\in U\,\, ,
\end{equation}
 independent of $t\geq 0$, where $ \phiL'(t,x)$ denotes the matrix of first order derivatives 
 of the components of $ \phiL(t,x)$. 
A necessary condition for
$\phiL$ to be non-contracting
is that $\Lambda_0$ is non-contracting, but in addition we need that the projection $\pi :\Lambda_t\to \R^n$ 
is not  singular. This implies that we have to stay away from caustics.

\subsection{The transport operator}
\label{subsec:transport}

Another description of the map $\phiL (t,x)$ is as follows: take the
trajectory 
$(p(t), q(t))=\Phi^t(\xi,x)$ which 
starts at $t=0$ at $q(t=0)=x$ with  initial momentum  $\xi=\nabla
S(x)$, then the 
map $\phiL(t,x)$ is the 
$q$ component of this trajectory, $\phiL(t,x)=q(t)$. Since by Hamilton's equations  $\dot q=\nabla_p H(p,q)$ and 
on $\Lambda_t$ we have $p=\nabla S(t,q)$, we find that $\phiL(t,x)$ is the unique solution to 
\begin{equation}\label{eq:flow_eq_F}
\frac{\ud \phiL(t,x)}{\ud t}=\nabla_pH(\nabla S(t,\phiL(t,x)),\phiL(t, x))\,\, ,\quad \text{with}\quad \phiL(0,x)=x\,\, .
\end{equation}
Notice that we allowed here for a general Hamiltonian, and not only one of the
simple form kinetic plus potential energy. 

We will now show that the transport operator $T(t)$ defined by \eqref{eq:defT} can be written  using  this map 
as 
\begin{equation}\label{eq:altDefT}
(T(t)A)(x)=[\det \phiL'(t,\phiL^{-1}(t,x))]^{-1/2} A(\phiL^{-1}(t,x))\,\, . 
\end{equation}
By comparing this formula with  \ref{app:vf} and using \eqref{eq:flow_eq_F} we see that the operator 
in \eqref{eq:altDefT} satisfies 
\begin{equation}
\ui\hbar \pa_t T(t)=\hat K(t) T(t)\,\, ,
\end{equation} 
where $\hat K(t)$ is the Weyl quantization of the function
\begin{equation}
K(t;p,q)=\nabla_pH(\nabla S(t,q),q)\cdot p\,\, .
\end{equation}
In particular we see that for the special case $H=\frac{1}{2}p^2+V(q)$ 
we have by \ref{app:vf}
\begin{equation}\label{eq:hatK}
\hat{K}(t)=\frac{\hbar}{\ui}\bigg[\nabla S(t,x)\nabla+\frac{1}{2}\Delta S(t,x) \bigg]
\end{equation}
and comparing with the transport equation \eqref{eq:trans_hbar} 
proves our claim.

\subsection{The metaplectic correction}

We can use this observation to give a more detailed description of the operator 
\begin{equation}\label{eq:delta_t}
-\frac{\hbar^2}{2}\Delta(t)=-\frac{\hbar^2}{2} T^*(t)\Delta T(t)\,\, .
\end{equation}
 This is an application of 
Egorov's theorem and the technical details are given in \ref{app:Eg}, where we show that  the Weyl symbol 
of this operator is given by 
\begin{equation}\label{eq:symb_deltat}
\delta H(t,\xi,x)=\frac{1}{2}\xi \cdot \cA(t,x) \xi +\frac{\hbar^2}{16}\sum_i \Tr [(\phiL^{-1})_i''(t,x)]^2\,\, ,
\end{equation}
where $\cA(t,x)$ is a symmetric matrix given by 
\begin{equation}
 \cA(t,x)=\big(\phiL'(t,x) \phiL'(t,x)^{\dagger}\big)^{-1}\,\, ,
 \end{equation}
and $(\phiL^{-1})_i''$ is the matrix of second derivatives of the $i$'th component of $\phiL^{-1}$. 
Bounds on the coefficients of the operator $\Delta(t)$ play an important role in estimating the 
accuracy of the time dependent WKB propagation and the metaplectic extension, because 
$\Delta(t)$ appears in the error terms \eqref{eq:D-I_est} and   \eqref {eq:error_meta}.
This is where the non-contraction condition we introduced above becomes important, 
by \eqref{eq:F_non_contr} we have that 
if $\phiL(t)$ is non-contracting then
\begin{equation}\label{eq:A_bounded}
\norm{\cA(t,x)}\leq C\,\, ,
\end{equation}
hence the operator $\Delta(t)$ has bounded coefficients. We can now give as well a more explicit description of the operator 
$\Delta_q(t)$. A short calculation shows that for a phase space function $H(\xi,x)$ which is quadratic in the momentum $\xi$
we have $\hbar L_q^*\hat{H}L_q =\hat H_q$ with $H_q(\xi, x)=H(\xi, q+\sqrt{\hbar}\, x)$ and so $ H_q(\xi, x)=H_q(\xi, q)+O(\sqrt\hbar)$. 
Applying this to $\delta H(t,\xi,x)=\frac{1}{2}\xi \cA(t,x) \xi+O(\hbar^2)$ gives that 
\begin{equation}\label{eq:delta_q_expl}
\Delta_q(t)=\nabla_x\cdot \cA(t,q)\nabla_x\,\, ,\quad \text{and}\quad \hbar L^*_q\Delta(t)L_q=\Delta_q(t)+O_t(\sqrt{\hbar})\,\, .
\end{equation}
If  $\phiL(t)$ is non-contracting then the remainder is  bounded uniformly in time. 

Since we have now an explicit expression for $\Delta_q(t)$ we can compute the action of the metaplectic 
operator $\m_q(t)$, defined by \eqref{eq:def_meta}, on a function $a$. Set 
\begin{equation}
C_t:=\int_0^t  \cA(s,q)\, \ud s
\end{equation}
and let $\hat{a}(\xi)=\int \ue^{-\ui x\xi} a(x)\, \ud x$ be the 
Fourier transform of $a$, then 
\begin{equation}
(\m_q(t)a)(x)=\frac{1}{(2\pi)^n}\int \ue^{-\frac{\ui}{2} \xi\cdot  C_t\xi}\hat{a}(\xi) \ue^{\ui x\cdot \xi}\, \ud \xi\,\, .
\end{equation}
Similarly we find for the rescaled operator $ M_q(t)$ that integrating \eqref{eq:def_meta_tilde} gives 
\begin{equation}\label{eq:intM}
(M_q(t)A)(x)=\frac{1}{(2\pi\hbar)^n}\int \ue^{-\frac{\ui}{\hbar }\frac{1}{2} \xi\cdot  C_t\xi}\hat{A}_{\hbar}(\xi)\ue^{\frac{\ui}{\hbar} x\cdot \xi}\, \ud \xi\,\, ,
\end{equation}
where $\hat A_{\hbar}(\xi)=\int \ue^{-\frac{\ui}{\hbar} x\xi} A(x)\, \ud x$. The operators $\m_q(t)$ and $M_q(t)$ act as Fourier multipliers 
with Gaussian functions. For an initial amplitude of the form $A(x)=(\pi\hbar)^{-n/4}\ue^{-\abs{x}^2/(2\hbar)}$ one finds in particular 
\begin{equation}
M_q(t)A(x)=\frac{1}{(\pi\hbar)^{n/4}}\frac{1}{\sqrt{\det(I+\ui C_t)}}\, \ue^{-\frac{1}{2\hbar}x\cdot(I+\ui C_t)^{-1}x}\,\, .
\end{equation}

 \subsection{The role of the initial manifold}\label{sec:initial_man}

The localised initial states we consider are of the form $\psi=(L_q a)\ue^{\frac{\ui}{\hbar} S}$, and the phase function 
determines the initial Lagrangian manifold which is crucial for the semiclassical propagation scheme we presented. 
The question we want to consider now is if the state $\psi$ determines the Lagrangian manifold uniquely, or in other words, 
if there might be other functions $\tilde a $ and $\tilde S$ such that $\psi=(L_q\tilde a)\ue^{\frac{\ui}{\hbar}\tilde S}$. 
In that case we would have some freedom in the choice of the initial Lagrangian manifold we use 
for propagation. 

The condition $(L_q a)\ue^{\frac{\ui}{\hbar} S}=(L_q\tilde a)\ue^{\frac{\ui}{\hbar}\tilde S}$ gives after multiplication by 
$\ue^{-\frac{\ui}{\hbar}\tilde S}$ and application of $L_q^*$ that 
\begin{equation}
\tilde a(x)=a(x)\ue^{\frac{\ui}{\hbar}[S(q+\sqrt\hbar\, x)-\tilde S(q+\sqrt\hbar\, x) ]}
\end{equation}
and by Taylor expansion we see that 
\begin{equation}
\frac{\ui}{\hbar}[S(q+\sqrt\hbar\, x)-\tilde S(q+\sqrt\hbar\, x)]=\frac{\ui}{\hbar}[S(q)-\tilde S(q)]+\frac{\ui}{\sqrt{\hbar}}[\nabla S(q)-\nabla\tilde S(q)]x
+\ui R(\hbar, q,x)\,\, , 
\end{equation}
where $R(\hbar, q,x)$ is a smooth real valued function.  The first term on the right hand side gives just a constant phase factor, so if 
\begin{equation}
\nabla S(q)=\nabla \tilde S(q)
\end{equation}
then $\tilde a$ defines a nice smooth function. But this conditions means that the two Lagrangian submanifolds 
$\Lambda$ and $\tilde \Lambda$, generated by $S$ and $\tilde S$, respectively,  intersect at $(p,q)=(\nabla S(q),q)$. 
We therefore conclude that in order to propagate a state localised around a phase space point 
$(p,q)$ we can use any Lagrangian manifold through that point which is non-contracting. 
This freedom of choice can be used to select  for instance initial manifolds for which the propagation 
becomes particularly easy, e.g., some for which no caustics develop.
 
In the case that the trajectory through $(p,q)$ is hyperbolic, any Lagrangian submanifold which is transversal 
to the stable directions is non-contracting, and hence can be used for propagation. Although it 
seems most natural to take the unstable manifold, any other manifold which is sufficiently transversal to stable direction will converge exponentially fast to the unstable manifold and hence 
should work with comparable efficiency.  We test this with the example in Section \ref{sec:exII}.

\subsection{Relation to standard coherent state propagation and time scales}

For times which are short compared to the Ehrenfest time our approach should reproduce the 
standard results on coherent state propagation. For the general class of initial amplitudes of the form \eqref{eq:scalingOp} 
the propagation is reviewed in  \cite{Pau97}.  A propagated coherent state is of the form 
\begin{equation}\label{eq:class_ch_state}
\psi(t,x)=\ue^{\frac{\ui}{\hbar}l(t)} \frac{1}{\hbar^{n/4}} a_{\hbar}\bigg(t,\frac{x-q(t)}{\sqrt{\hbar}}\bigg)\ue^{\frac{\ui}{\hbar} p(t)\cdot (x-q(t)}
\end{equation}
where $(p(t),q(t))$ is a classical trajectory, $l(t)$ an action type phase which includes as well Maslov terms, and 
$a_{\hbar}(t,x)=a_0(t,x)+\sqrt{\hbar}\, a_{1}(t,x)+\cdots$ with leading term given as the solution to
\begin{equation}\label{eq:coh_state_meta}
\ui\pa_ta_0(t,x)=\bigg[-\frac{1}{2}\Delta + \frac{1}{2} x\cdot V''(q(t)) x\bigg]a_0(t,x)\,\, ,\quad a_0(0,x)=a_0\,\, .
\end{equation}
This means that the centre of the state is propagated along the trajectory $(p(t),q(t))$ and its shape changes
according to the linearised dynamics around the centre.

If we expand $S(t,x)$  around 
$x=q(t)$ up to second order then we find that we can write our approximation \eqref{eq:TD_approx} in the form 
\eqref{eq:class_ch_state} with 
\begin{equation}
a_{\hbar}(t,x)= b(t,x)\ue^{\frac{\ui}{2}x\cdot S''(t,q(t))x+O(\sqrt{\hbar})}\,\, ,\,\, \text{where}\quad b(t,x)=[L^*_{q(t)} T(t)L_qD_q(t)a](x)
\end{equation}
Now using the operator identities 
\begin{align}
\ui \pa_t L^*_{q(t)}&= L^*_{q(t)}\ui \pa_t+L^*_{q(t)} \ui\dot q\cdot \nabla \,\, , \\
\sqrt{\hbar}L^*_{q(t)}\nabla L_{q(t)}&=\nabla\,\,  , \\
L^*_{q(t)} \nabla S(t,x)L_{q(t)}&=\nabla S(t,q(t)) +\sqrt{\hbar}\,   S''(t, q(t)) x+O(\hbar)\,\, ,\\
L_{q(t)}^*T(t)L_q \Delta_qL^*_qT^*(t)L_{q(t)}&=\Delta +O(\sqrt{\hbar})\,\, ,
\end{align}
which all follow by direct computation, 
we obtain with $\nabla S(t,q(t))=p(t)=\dot q(t)$ that 
\begin{equation}\label{eq:b_der}
\ui\pa_t b(t,x)=-\bigg(\ui x\cdot S''\nabla +\frac{\ui}{2}\Delta S +\frac{1}{2}\Delta\bigg) b(t,x)+O(\sqrt{\hbar})\,\, , 
\end{equation}
where $S$ and its derivatives are all evaluated at $x=q(t)$. If we furthermore use 
$\Delta( b(t,x)\ue^{\frac{\ui}{2}x\cdot S''x})=[\Delta b(t,x)+2\ui \nabla b(t,x)\cdot S'' x +
(-\abs{S''x}^2+\ui\Delta S)b(t,x)]\ue^{\frac{\ui}{2}x\cdot S''x}$ and combine this with \eqref{eq:b_der}  we find
\begin{equation}
\ui\pa_t a_h(t,x)=\bigg[-\frac{1}{2}\Delta -\frac{1}{2}x\cdot \dot S'' x-\frac{1}{2}\abs{S'' x}^2\bigg]a_h(t,x)+O(\sqrt{\hbar})\,\, .
\end{equation}
Finally from the Hamilton-Jacobi equation \eqref{eq:HamJac} we get $-x\cdot \dot S'' x-\abs{S'' x}^2/2=x\cdot V'' x$, hence 
the leading term $a_0(t,x)$ satisfies \eqref{eq:coh_state_meta}. 

This was a formal computation which showed that our result reproduces the previously existing  results, but it 
didn't gave much insight where the  standard approximation might break down. To see this more clearly let us 
just look at the amplitude $T(t)L_q b$ with $b=D_qa$. By Taylor expansion around $x=q(t)=\phiL(t,q)$ we have 
$\phiL^{-1}(x)-q\approx [\phiL'(t,q(t))]^{-1}(x-q(t))$ and hence 
\begin{equation}
T(t)L_q b(x)\approx\frac{1}{\hbar^{\frac{n}{4}}\sqrt{\det \phiL'(t,q(t))}} \, b\big( [\sqrt{\hbar}\, \phiL'(t,q(t))]^{-1}(x-q(t))\big)\,\, .
\end{equation}
From this expression we see that the state will be localised around $x=q(t)$ if 
\begin{equation}
\sqrt{\hbar}\norm{\phiL'(t,q(t))}\ll 1\,\, , 
\end{equation} 
and since $\phiL(t,x)$ is a projection of the flow from phase space this is satisfied if  $t\ll T_E(\hbar)$. On the other hand side, if 
$\sqrt{\hbar}\norm{\phiL'(t,q(t))}\approx 1$, then the amplitude is no longer localised around a point, and the state 
becomes a Lagrangian, or WKB state.


\section{A generalised propagation scheme and caustics} 
\label{sec:gen_scheme}

So far we have avoided the discussion of caustics, but although there are situations where no caustics occur, in many 
interesting situations we expect caustics to develop.  
 A caustic is the image in  position space $\R^n$ of the singularities of  the projection $\pi_{\Lambda_t}:\Lambda_t\to \R^n$, 
 that means at a caustic  the tangent plane $T_z\Lambda_t$  to $\Lambda_t$ at a point $z=(p,q)\in \Lambda_t$ becomes vertical in the sense 
that $T_z\Lambda_t\cap V_z\neq \{0\}$ where $V_z=\{ (\xi,q)\,\, ;\, \xi\in\R^n\}$ denotes the vertical subspace. In a neighbourhood of a caustic the manifold 
$\Lambda_t$ can no longer be represented by a  generating function $S(t,x)$ as in \eqref{eq:genLt}, and 
so the simple time dependent WKB propagation we presented above can no longer work. 
This problem is usually solved by switching to a different representation of the quantum state, e.g., taking the Fourier transform 
switches from the position representation to the momentum representation and in this new representation we have to consider 
instead  the projection of $\Lambda_t$ to momentum space.

We want to use a method which allows for a bit flexibility, so that we can accommodate all possible different representations. 
To this end we look at how we can in general approximate the full Hamiltonian $H$ by  a simpler Hamiltonian $H_1$ in a way 
that the propagation of Lagrangian states initially  associated with $\Lambda$ can be approximated using $H_1$. 
The crucial observation which will guide our conditions on $H_1$ is that if $B(p,q)$ and $A(x)$ are  sufficiently smooth functions and 
$\hat B$ is the Weyl quantisation of $B$, then 
\begin{equation}
\hat B\big(A\ue^{\frac{\ui}{\hbar}S}\big)=\big(B_{\Lambda}A+O(\hbar)\big)  \ue^{\frac{\ui}{\hbar}S}
\end{equation}
where $B_{\Lambda}(x)=B(\nabla S(x),x)$, this is a classical result, see, e.g, \cite{Dui11}.      This means that the Lagrangian state is 
concentrated in phase space on $\Lambda$, in particular if $B_{\Lambda}=0$, then
$\hat B \psi=O(\hbar)$. More generally, if $B$ vanishes of order $N$ on $\Lambda$, i.e., 
$(\pa^{\alpha}B)|_{\Lambda}=0$ for $\abs{\alpha}\leq N$, then 
\begin{equation}\label{eq:vanishLambda}
\hat B\psi=O(\hbar^N)\,\, .
\end{equation}

Since a propagated Lagrangian state is concentrated near the propagated Lagrangian manifold 
$\Lambda_t=\Phi^t(\Lambda)$, where $\Phi^t$ is the Hamiltonian flow generated by $H$, we expect that if 
$H_1$ is close to $H$ near $\Lambda_t$ then it will provide an accurate approximation for the purpose of 
propagating states concentrated near $\Lambda_t$. 
 The  conditions for a function  $H_1$  to be a general first 
order approximation of $H$  near 
 $\Lambda_t$ are 
\begin{equation}\label{eq:HH1approx}
(H-H_1(t))|_{\Lambda_t}=0\,\, ,\quad\text{and}\quad (\nabla H-\nabla H_1(t))|_{\Lambda_t}=0 \,\, .
\end{equation}
These conditions ensure that the Hamiltonian vector fields generated by $H$ and $H_1$ agree on $\Lambda_t$, 
hence if we denote by $\Phi^t_1$ the time $t$ map generated by $H_1$, then 
\begin{equation}\label{eq:eq_class}
\Phi^t|_{\Lambda}=\Phi_1^t|_{\Lambda}\,\, ,
\end{equation}
i.e., the classical dynamics agree when restricted to $\Lambda$. 

Let us now turn to the quantisations of $H$  and $H_1$. Let $U(t)$ and $U_1(t)$ be the time evolution operators generated by 
$\hat H$ and $\hat H_1$, respectively, then we expect by \eqref{eq:eq_class} that $U(t)\psi \approx U_1(t)\psi$  for a 
Lagrangian state associated with $\Lambda$.  To verify this we make an Ansatz 
\begin{equation}
U(t)=U_1(t)V(t)\,\, , 
\end{equation}
 and inserting this into the Schr{\"o}dinger equation for $U(t)$ gives for $V(t)$ the equation
\begin{equation}
\ui\hbar \pa_t V(t)=\widehat{\delta H}(t) V(t)\,\, ,\quad \text{with}\quad V(0)=I\,\, ,
\end{equation}
where 
\begin{equation}
\widehat{\delta H}(t)=U_1^*(t)[\hat H-\hat H_1(t)]U_1(t)\,\, . 
\end{equation}
By Egorov's theorem, see \ref{app:egorov},  the operator 
$\widehat{\delta H}(t)$ is in leading order in $\hbar$ the quantisation of $H-H_1$ transported along the map $\Phi_1^t$, 
i.e.,
\begin{equation}
\delta H(t)=[H-H_1(t)]\circ \Phi_1^t+O_t(\hbar^2)\,\, .
\end{equation}
This implies by \eqref{eq:HH1approx} that 
\begin{equation}
\delta H(t)|_{\Lambda_0}=O_t(\hbar^2) \,\, ,\quad \text{and}\quad \nabla \delta H(t)|_{\Lambda_0}=O_t(\hbar^2)\,\, ,
\end{equation}
and therefore by \eqref{eq:vanishLambda} we have  for an initial 
state of the form $\psi=A_0\ue^{\frac{\ui}{\hbar}S_0}$, where $A_0$ has bounded derivatives, that 
\begin{equation}\label{eq:deltaHsmall}
\norm{\widehat{\delta H}(t)\psi}=O_t(\hbar^2)\,\, .
\end{equation}
Thus  the generator of $V(t)$ is small when applied to $\psi$ and therefore  Duhamel's principle, \eqref{eq:Duhamel}, gives 
\begin{equation}
 V(t)\psi=\psi+O_t(\hbar )\,\, ,
 \end{equation}
hence $U_1(t)\psi$ is a good approximation for $U(t)\psi$ if $\psi$ is a Lagrangian state.

This is a generalisation of the standard time dependent WKB method presented in Section \ref{sec:realTDWKB}. 
The choice of $H_1$ corresponding to the results in  Section \ref{sec:realTDWKB} is 
\begin{equation}\label{eq:H1approxP}
H_1(t;\xi,x):=H(\nabla S(t,x),x)+\nabla_{\xi}H(\nabla S(t,x),x)\cdot \xi
\end{equation}
which we obtain from $H(p,q)=H(\nabla S(t,x)+\xi,x)$ as a first order approximation in the shifted momentum 
$\xi=p-\nabla S(t,x)$. For a Hamiltonian of the form $H=\frac{1}{2}p^2+V(q)$ one finds using 
\ref{app:vf} that 
$\hat H_1=H(\nabla S(t,x),x)-\ui\hbar \nabla S(t,x)\cdot \nabla -\frac{\ui \hbar}{2}\Delta S(t,x)$, and so 
by comparing this with \eqref{eq:defT} and \eqref{eq:HamJac} we see that in this case we have 
\begin{equation}\label{eq:U1TDV}
U_1=\ue^{\frac{\ui}{\hbar}S(t)}T(t)\ue^{-\frac{\ui}{\hbar}S_0}\,\, \quad\text{and}\quad V(t)=\ue^{\frac{\ui}{\hbar}S_0}D(t)\ue^{-\frac{\ui}{\hbar}S_0}\,\, .
\end{equation}
So how does a caustic affect this relation? In defining $H_1$ we used coordinates $(\xi,x)$ near $\Lambda_t$ defined by
$p=\nabla S(t,x)+\xi$ and $q=x$, so that $\Lambda_t=\{\xi=0\}$, but at a caustic this coordinate system becomes singular. 
In order to repair this we can  just use a different canonical coordinate system near $\Lambda_t$, one which does not 
become singular, and choose for $H_1$ the first order Taylor approximation in the  transversal coordinates.

The only problem we might encounter with this more general version of time dependent WKB theory is that 
the control of the time dependence of the remainder term \eqref{eq:deltaHsmall} becomes more involved. 
In Section \ref{sec:realTDWKB} we were able to take advantage of the very explicit version of Egorov's theorem 
from \ref{app:Eg} to reduce this problem to the non-contractiveness of the transport map $\phiL(t)$. For different 
choices of $H_1$ this problem can become much harder and we reserve a more thorough investigation 
for the future. But this question concerns our ability to get explicit error estimates, it does not prevent us 
to use, after suitable testing, the method for explicit numerical propagation in concrete problems.

Now if the initial state $\psi_0$ is localised in phase space around a point $(p,q)\in \Lambda_0$, so that the estimate 
\eqref{eq:deltaHsmall} no longer holds,  then it is natural 
to approximate $V(t)$ by taking only the behaviour of its generator $\delta H(t,\xi,x)$ near $(p,q)$ into account. 
By \eqref{eq:HH1approx} we have $\delta H(t,p,q)=0$ and $\nabla\delta H(t,p,q)=0$, hence  the simplest nontrivial 
approximation is quadratic, with $z=(\xi,x)$ and $z_0=(p,q)$ we have 
\begin{equation}
\delta H(t,x,\xi)\approx \delta H_2(t,z)=\frac{1}{2}(z-z_0)\delta H''(t,z_0)(z-z_0)\,\, .
\end{equation}
The time evolution $\tilde M_{p,q}(t)$ generated by $\widehat{\delta H}_2(t)$ via 
$\ui\hbar \pa_t  M_{p,q}(t)=\widehat{\delta H}_2(t)  M_{p,q}(t)$ and 
$ M_{p,q}(0)=I$ is  a metaplectic operator since the generator 
$\widehat{\delta H}(t)$ is the quantisation of a quadratic function on phase space. 
Therefore for $\psi$ strongly localised around $(p,q)$ we expect 
$V(t)\psi\approx   M_{p,q}\psi$ and hence we have arrived now at the more general version of 
\eqref{eq:TM_approx}, which reads
\begin{equation}
U(t)\psi= U_1(t) M_{p,q}(t)\psi+O_t(\sqrt{\hbar})
\end{equation}
if $\psi$ is concentrated in phase space at $(p,q)$ as in \eqref{eq:scalingOp}. 

The relation of $ M_{p,q}(t)$ to $ M_q(t)$ follows from  \eqref{eq:U1TDV}, if $H_1$ is of the form 
\eqref{eq:H1approxP} we find that 
 \begin{equation}
  M_{p,q}(t)=\ue^{\frac{\ui}{\hbar} S_0^{(2)}}  M_q(t)\ue^{-\frac{\ui}{\hbar} S_0^{(2)}}
 \end{equation}
 where $p= \nabla S_0(q)$ and $S_0^{(2)}(x)=p(x-q)+\frac{1}{2}(x-q)S_0''(q)(x-q)$ is the quadratic part 
 of the Taylor expansion of $S_0$ around $x=q$.

Since $ M_{p,q}(t)$ has a quadratic generator it is the quantisation of a linear symplectic map 
 $P_{p,q}(t)$ on $T_{p,q}(\R^n\times \R^n)$ and by construction this map can be expressed 
 in terms of the linearisations of the maps $\Phi^t$ and $\Phi^t_1$. Since 
 $V(t)=U_1^{-1}(t)U(t)$ we can view $V(t)$ as a quantisation of 
 the map $(\Phi_1^{t})^{-1}\circ\Phi^t$, and $ M_{p,q}(t)$ is then the quantisation of the 
 linearisation of that map around $p,q$, i.e., 
 \begin{equation}\label{eq:PrelPhi}
 P_{p,q}(t)=(\ud \Phi_1^{t})^{-1}\ud \Phi^t\,\, .
 \end{equation}
 Since  $\Phi^t$ and $\Phi_1^t$ are identical on $\Lambda$ we have that
 $P_{p,q}(t)|_{T_{p,q}\Lambda_0}=I$, hence $P_{p,q}$ is a shear relative to the tangent space $T_{p,q}\Lambda_0$ of the 
 initial Lagrangian manifold at $(p,q)$.

 In case that $H_1$ is given by \eqref{eq:H1approxP} the map $P_{p,q}(t)$ can be described in some more detail. Note that 
 since $H_1$ in \eqref{eq:H1approxP} contains only a linear term in the momentum we find that Hamilton's equation for $H_1$ give  
 \begin{equation}
 \dot x=\nabla_pH(\nabla S(t,x),x)\,\, , 
 \end{equation}
that means the trajectory of $x$ under $\Phi_1^t$ does not depend on the initial momentum $\xi$. Hence  
 $\Phi^t_1$ maps vertical subspaces into 
vertical subspaces, i.e., $\ud \Phi^t_1 (V_{z_0})\subset  V_{z(t)}$, where 
for $z=(p,q)$ we set $V_z=\{(\xi,q)\, ,\xi\in\R^n\}$. Therefore the map $P_{p,q}(t)$ satisfies 
\begin{equation}
P_{p,q}(t)^{-1}|_{T_{p,q}\Lambda_0}=I\quad \text{and}\quad P_{p,q}(t)^{-1}(V_{(p,q)})=(\ud \Phi^t)^{-1}(V_{\Phi^t(p,q)})
\end{equation}
and as we show in \ref{app:symp_lemma}  this implies that  the knowledge of the Lagrangian subspace $(\ud \Phi^t)^{-1}(V_{\Phi^t(p,q)})$ determines the map $P_{p,q}(t)^{-1}$ and hence   $P_{p,q}(t)$ uniquely. 

Let us look at two examples where 
we can determine the long time limit of 
$(\ud \Phi^t)^{-1}(V_{\Phi^t(p,q)})$ from dynamical conditions.

\begin{itemize}
\item[(a)] If we have a one-dimensional system and $\pa^2 H/\pa p^2>0$, then the velocity increases with $p$ and hence 
$ (\ud \Phi^t)^{-1}(V_{\Phi^t(p,q)})\to \{(0,q), q\in \R\}$ for $t\to\infty$. Therefore if $T_{(p,q)}\Lambda_0\neq  \{(0,q), q\in \R\}$ 
we can use the result from  \ref{app:symp_lemma} and see that for large $t$ the map $P_{p,q}(t)$ tends to a limit 
 $P_{p,q}^{\infty}$ and therefore the corresponding metaplectic correction $ M_{p,q}(t)$ tends to a limit, too. 
\item[(b)] If the trajectory through $(p,q)$ is hyperbolic and the vertical subspaces $V_{z}$ are transversal to the 
unstable subspaces $V^{u}_z$ along the trajectory, then $ (\ud \Phi^t)^{-1}(V_{\Phi^t(p,q)})$ tends to the stable subspace 
$V_{p,q}^{s}$ for large $t$, and at an exponential rate if the hyperbolicity is uniform. So in this situation we find that for large $t$ 
the map $P_{p,q}(t)$ tends to a limit 
 $P_{p,q}^{\infty}$ which is uniquely defined by the conditions that $P_{p,q}^{\infty}|_{T_{p,q}\Lambda_0}=I$ and 
 $P_{p,q}^{\infty}(V^s_{p,q})=V_{p,q}$.  Therefore in this situation the metaplectic correction tends for large $t$ to a limit, too, 
 and expontially fast if the hyperbolicity is uniform. 
\end{itemize}


\section{Examples}
\label{sec:exI}

In order to illustrate how the theory we developed in the previous sections works, it 
is very instructive to look at a couple of simple examples. They will allow us 
to understand in more detail the interplay among the classical dynamics,  the choice of an initial Lagrangian manifold and 
the analytical constructions we developed.

\subsection{The free particle}

The first example we look at is the free particle in 1 dimension, with Hamilton operator 
$-\frac{\hbar^2}{2}\Delta$. We want to propagate a state 
that is initially localised at $(p,q)\in\R^2$, where $p\geq 0$. As initial 
phase function we choose 
\begin{equation}\label{eq:S_0}
S_0(x)=p(x-q)+\frac{\alpha}{2}(x-q)^2
\end{equation}
where $\alpha\in \R$ is a real parameter. The corresponding 
Lagrangian manifold is 
\begin{equation}
\Lambda_0=\{(p+\alpha x, q+x)\, ,\, x\in\R\}
\end{equation}
which is a line through $(p,q)$ with slope $\alpha$.  The corresponding classical Hamilton function
is $H(\xi)=\frac{1}{2}\xi^2$ and the Hamiltonian flow it generates is given by 
$\Phi^t(\xi,x)=(\xi,x+t\xi)$. Applying the flow to $\Lambda_0$ gives 
$\Lambda_t=\Phi^t(\Lambda_0)=\{(p+\alpha x, q+x +t(p+\alpha x))\, ,\, x\in \R\}$ and by replacing $x$ by $x/(1+\alpha t)$ this can be rewritten as 
\begin{equation}
\Lambda_t=\{(p+\alpha(t) x, q(t)+x)\, ,\, x\in \R\}\, \, , \quad \text{where}\quad \alpha(t)=\frac{\alpha}{1+\alpha t}\,\, \, ,\, \, q(t)=q+tp\,\, ,
\end{equation}
which is a line through $(p,q(t))=\Phi^t(p,q)$ with slope $\alpha(t)$. Notice that if $\alpha<0$, i.e., if the initial line has negative slope, 
then $\alpha(t)\to\infty$ for $t\to -1/\alpha$, this means that the line $\Lambda_t$ turns vertical, hence we have a caustic. 
But if $\alpha \geq 0$, then $\alpha(t)\leq \alpha$ for all $t\geq 0$, and no caustics occur. The phase function 
which generates $\Lambda_t$ and satisfies the Hamilton-Jacobi equation with initial condition $S_0$ is given by 
\begin{equation}
S(t,x)=\frac{1}{2} p^2t+p(x-q(t))+\frac{\alpha(t)}{2}(x-q(t))^2\,\, .
\end{equation}

To find the transport operator we have to find the map $\phiL(t,x)=\pi_{\Lambda_t}\Phi^t(\pi^{-1}_{\Lambda_0}(x))$.
To this end we notice  that $\pi^{-1}_{\Lambda_0}(x)=(\nabla S_0(x),x)=(p+\alpha (x-q),x)$, hence 
$\Phi^t(\pi^{-1}_{\Lambda_0}(x))=(p+\alpha (x-q),x+t(p+\alpha (x-q))$ and the final projection just takes the second component, 
therefore
\begin{equation}
\phiL(t,x)=(1+\alpha t)x+t(p-\alpha q)\,\, .
\end{equation}
Since $\phiL'(t,x)=(1+\alpha t)$ the condition $\alpha\geq 0$ guarantees that $\phiL$ is 
non-contracting and we get  $\phiL^{-1}(t,x)=\frac{1}{1+\alpha t} \big(x-t(p-\alpha q)\big)$. Therefore the action 
of the transport operator reads 
\begin{equation}
T(t)A(x)=\frac{1}{(1+\alpha t)^{1/2}}A\bigg(\frac{1}{1+\alpha t} \big(x-t(p-\alpha q)\big)\bigg)\,\, ,
\end{equation}
which for $\alpha\geq 0$ is well defined for all $t\geq 0$. In order to compute the metaplectic correction, which is generated by 
\eqref{eq:symb_deltat}, we notice that 
$\cA(t,x)=(1+\alpha t)^{-2}$ is independent of $x$, hence we have 
\begin{equation}
\Delta_q(t)=\Delta(t)=\frac{1}{(1+\alpha t)^2}\Delta\,\, .
\end{equation}
Then  the metaplectic correction \eqref{eq:intM} is 
\begin{equation}
 M_q(t)=\ue^{\frac{\ui\hbar }{2} \frac{t}{1+\alpha t}\Delta}\,\, ,
\end{equation}
where the time dependent factor in the exponent comes from the integration of the time dependence in $\Delta(t)$, 
$\int_0^t\frac{1}{(1+\alpha t')^2}\, \ud t'=\frac{t}{1+\alpha t}$. We see that in particular, as we observed at the end of 
Section \ref{sec:gen_scheme}, that the operator tends to a limit for large $t$, 
$ M_q(t)\to M_q(\infty)=\ue^{\frac{\ui\hbar }{2} \frac{1}{\alpha }\Delta}$.

We computed all the elements in our extended time dependent WKB propagation scheme for the free particle. 
Let us apply this to an initial state $\psi_0(x)=A_0(x)\ue^{\frac{\ui}{\hbar}[p(x-q)+\frac{\alpha}{2}(x-q)^2]}$ with $A_0=L_qa_0$ for some $a\in S(\R)$. Then 
$M_q(t)A_0=L_qa_t$ with $a_t=\m_q(t)a_0$ and since $\m_q(t)$ tends to a limit for large $t$, we have that $a_t\in S(\R)$ with bounds uniform in $t\in\R^+$.   Hence we find 
\begin{equation}\label{eq:free_ev_state}
\psi(t,x)=A(t,x) \ue^{\frac{\ui}{\hbar}[\frac{1}{2} p^2t+p(x-q(t))+\frac{\alpha(t)}{2}(x-q(t))^2]}\,\, ,
\end{equation}
with
\begin{equation}\label{eq:free_ev_amplitude}
A(t,x)=T(t)M_q(t)A_0(x)=\frac{1}{\hbar^{1/4}(1+\alpha t)^{1/2}}a_t\bigg(\frac{1}{\hbar^{1/2}(1+\alpha t)} \big(x-t(p-\alpha q)\big)\bigg)\,\, .
\end{equation}
Since the Hamiltonian is quadratic this expression is actually exact, the remainder terms are zero. 
Notice that the behaviour of the amplitude $A(t,x)$ depends on the size of $\hbar^{1/2}(1+\alpha t)$, which determines the Ehrenfest time scale 
\eqref{eq:Ehrenfest_integrable}.    If $\hbar^{1/2}(1+\alpha t)\to 0$ 
the amplitude becomes concentrated, but if $\hbar^{1/2}(1+\alpha t)\sim 1$  then  the amplitude $A(t,x)$ is a smooth function and 
$\psi(t)$ becomes actually a Lagrangian state.  

Let us note that for the special case that our initial state is a Gaussian coherent state, 
\begin{equation}
\psi_0(x)=\frac{1}{(\pi\hbar)^{\frac{1}{4}}} \ue^{\frac{\ui}{\hbar}[p(x-q)+\frac{b}{2}(x-q)^2]}\,\, ,
\end{equation}
 our result  is identical to the one obtained by
 the standard propagation of coherent states, since the Hamiltonian is quadratic. 
 The only difference is that we derived it in a different way. 
But it was still instructive to go through the derivation 
since it highlights a few important points: We can rewrite the state using the phase function $S_0$ from \eqref{eq:S_0}  as
\begin{equation}
\psi(x)=[L_qa_0](x)\ue^{\frac{\ui}{\hbar}S_0(x)} \,\, ,\quad\text{with}\quad a_0(x)=\ue^{\ui \frac{b-\alpha}{2}\, x^2}\,\, ,
\end{equation}
and then applying \eqref{eq:free_ev_state} and \eqref{eq:free_ev_amplitude} gives us the propagated state. 
But since the initial state is independent of $\alpha$, the final result is independent of the choice of $\alpha$, too, i.e., of the initial Lagrangian sub-manifold used to propagate the state. But the intermediate steps depend on $\alpha$, first of all, we needed to choose $\alpha\geq0$ 
in order to avoid caustics, then the choice of $\alpha$ determined how we split the evolution into transport, implemented 
by the operator  $T(t)$,  and dispersion, represented by $D(t)= M_0(t)$. The simplest choice would have  been
$\alpha=0$, then $\Lambda_0$ is horizontal, and actually invariant under the classical flow. 
In this case $T(t)$ is defined by  transport along one central trajectory, and all the dispersion is described by $D(t)$. If $\alpha>0$, then 
$\Lambda_0$ is transversal to the flow and  the action of $T(t)$ takes actually a family of trajectories into 
account. Since these trajectories move with different speed, this already accounts for part of the dispersion 
of the wavepacket, and the operator $M(t)$ has only to add the dispersion transversal to $\Lambda_t$. 
This is reflected by the fact that the coefficients of $\Delta(t)$ tend to $0$ for large $t$, and fast enough so that 
the limit 
\begin{equation}
\lim_{t\to\infty} M_0(t)=\ue^{\frac{\ui\hbar}{2}\frac{1}{\alpha}\Delta}
\end{equation}
actually exist if $\alpha>0$. This means that for large $t$ we can replace $ M_0(t)$ by its limit and still get a 
good approximation of the propagated state. This is why $\alpha>0$ is preferable to $\alpha=0$, 
in particular if we extend these construction to more general systems. 

In addition, the fact that for $\alpha>0$ the coefficients of $\Delta(t)$ tend to $0$ for large $t$ reflects the property that  the map 
$\phiL$ is expanding, this has as well consequences for the size of the remainder term if we 
add higher order terms to the Hamiltonian. To study this effect will be the main focus of the next example.

\subsection{Integrable systems}

Let us look now at a more general Hamiltonian which is a function of the momentum only, 
but not necessarily quadratic anymore, so $H=H(\xi)$, and we will assume that 
for $\xi>0$ we have $H'(\xi)>0$ and $H''(\xi)\geq 0$. This is the form a  Hamiltonian of an integrable system takes in 
action-angle coordinates. 

The classical dynamics is given by $\Phi^t(\xi,x)=(\xi, x+tH'(\xi))$ and by the 
condition $H''(\xi)\geq 0$ the velocity $H'(\xi)$ can not decrease with increasing $\xi$.
This implies that if we take again an initial phase function of the form  $S_0(x)=px+\alpha x^2/2$ then 
the corresponding Lagrangian submanifold does not develop caustics for $t\geq 0$ if $\alpha\geq 0$, 
We find $\Lambda_t=\Phi^t(\Lambda_0)=\{(p+\alpha x, x+tH'(p+\alpha x))\,\, ,x\in \R\}$ and therefore 
\begin{equation}\label{eq:phiL_int}
\phiL(t,x)=x+tH'(p+\alpha x)\, \, ,\quad \text{and}\quad \phiL'(t,x)=1+\alpha tH''(p+\alpha x)
\end{equation}
so if $\alpha>0$ and $H''>0$, then $\phiL(t,x)$ is actually expanding. Since the flow is now no longer 
linear,  $\Lambda_t$ is no longer a straight line and we do not attempt to find an explicit expression for 
$S(t,x)$, but rather take it as defined by the Hamilton-Jacobi equation. But we observe that since 
$\Lambda_t=\{(\nabla S(t, \phiL(t,x)), \phiL(t,x))\}$ we have  $\nabla S(t, \phiL(t,x))=p+\alpha x=\nabla S_0(x)$ which 
is of course a consequence of momentum conservation.

  Having the phase function and the map $\phiL(t)$, and therefore the transport 
operator $T(t)$, we have the ingredients of the standard WKB propagation, and now we want to compute the 
metaplectic correction for an initial state localised at $(p,0)$. Here we have to use the generalised formulation 
from Section \ref{sec:gen_scheme} since the Hamiltonian is no longer a sum of a quadratic kinetic energy term  and a potential. 
 The transport operator is the quantisation of the map 
$\Phi^t_1(\xi,x)=([\phiL'(t,x)^{\dagger}]^{-1}\xi , \phiL(t,x))$
and so the operator $\widehat{\delta H} $ has symbol $\delta H(t,\xi,x)=\delta H^{(0)}(t,\Phi^t_1(\xi,x))+O_t(\hbar^2)$ where 
 $\delta H^{(0)}(t,\xi,x)=H(\nabla S(t,x)+\xi)-H(\nabla S(t,x))-\xi\cdot \nabla H (\nabla S(t,x))$ and the $O_t(\hbar^2)$ term comes from Egorov's theorem. Therefore  
using $\nabla S(t,\phiL(t,x))=\nabla S_0(x)$ we find 
\begin{equation}
\begin{split}
\delta H(t,\xi, x)=&H(\nabla_0 S(x)+[\phiL'(t,x)^{\dagger}]^{-1}\xi)\\
&-H(\nabla S_0(x))-[\phiL'(t,x)^{\dagger}]^{-1}\xi\cdot \nabla H (\nabla S_0(x))+O_t(\hbar^2)\,\, .
\end{split}
\end{equation}
Since the point $(p,0)$ corresponds to $\xi=0, x=0$ we find using \eqref{eq:phiL_int} that 
the generator of the metaplectic 
correction is $\widehat{\delta H_2}(t)=-\frac{\hbar^2}{2}\frac{H''(p)}{(1+\alpha t H''(p))^2} \Delta$ and hence  
\begin{equation}
 M_0(t)=\ue^{\frac{\ui\hbar}{2}\frac{H''(p)t }{1+\alpha t H''(p)} \Delta}\,\, . 
\end{equation}
We observe the same phenomenon as in the free case, for $\alpha>0$ the large $t$ behaviour has 
a limit. But  now this is as well reflected in the behaviour of the remainder term which is 
determined by the size of 
\begin{equation}
\delta H(t,\sqrt{\hbar}\xi, \sqrt{\hbar}x)-\delta H_2(t,\sqrt{\hbar}\xi, \sqrt{\hbar}x)=
O\big(\hbar^{3/2}[\phiL'(t,x)\phiL'(t,x)^{\dagger}]^{-1}\big)\,\, ,
\end{equation}
 and since the time integral of this quantity is bounded we get  
 for a coherent state $\psi_0$ concentrated at 
$(p,0)$ that
\begin{equation}
\norm{U(t)\psi_0-U_1(t) M(t)\psi_0}\leq C\sqrt{\hbar}\,\, . 
\end{equation}
Here we made  the assumption that we can control as well the remainder term $O_t(\hbar^2)$ from the application of 
Egorov's theorem, which seems likely since the transport operator has still a simple form. But we leave a detailed investigation 
to a future publication.  

This is in contrast to the standard coherent state propagation which can only work if $\abs{t}\sqrt{\hbar}\ll1$, since 
$T_E=1/\sqrt{\hbar}$ for an integrable system. The physical reason for this is that for 
$t\sqrt\hbar\sim1 $ the state 
$U_1(t) M(t)\psi_0$ is no longer a coherent state  but has become a Lagrangian state, i.e., there is a qualitative change in the nature of the state 
at the Ehrenfest time.

\subsection{A parabolic barrier}

Let us now look at a case with a hyperbolic trajectory, let $H(\xi,x)=\frac{1}{2}\xi^2-\frac{V_0}{2}x^2$, this 
Hamiltonian describes the motion in a parabolic barrier if $V_0>0$. The classical flow is given by  
\begin{equation}
\Phi^t(\xi,x)=(\cosh(\lambda t)\xi+\lambda \sinh(\lambda t)x,\lambda^{-1}\sinh(\lambda t)\xi+\cosh(\lambda t)x)
\end{equation}
where $\lambda:=\sqrt{V_0}>0$ is the Liapunov exponent. The origin is a hyperbolic fixed point with stable and unstable 
manifolds given by 
\begin{equation}
V^{s}=\{(-\lambda x,x)\, ,\, x\in\R\}\,\, ,\quad V^{u}=\{(\lambda x,x)\, \, ,\, x\in\R^n\}\,\, .
\end{equation}

Let us first choose an initial state localised on the fixed point, i.e., at $(0,0)$. An initial phase function 
$S_0(x)=\alpha^2 x^2/2$ corresponds to $\Lambda_0=\{(\alpha x,x)\, ,\, x\in\R\}$ and 
the transported manifold $\Lambda_t=\Phi^t(\Lambda_0)$ can be written as 
\begin{equation}
\Lambda_t=\{(\alpha(t) x,x)\, ,\, x\in\R\}\,\, ,\quad\text{with}\quad\alpha(t)=\frac{\cosh(\lambda t)\alpha+\lambda \sinh(\lambda t)}{\lambda^{-1}\sinh(\lambda t)\alpha+\cosh(\lambda t)}\,\, . 
\end{equation}
The corresponding phase function is $S(t,x)=\alpha(t)x^2/2$ and we find furthermore 
\begin{equation}
\phiL(t,x)=(\alpha \lambda^{-1} \sinh(\lambda t)+\cosh(\lambda t))x\,\, ,
\end{equation}
and therefore
\begin{equation}
\Delta(t)=(\alpha \lambda^{-1} \sinh(\lambda t)+\cosh(\lambda t))^{-2}\Delta\,\, .
\end{equation}

The manifold $\Lambda_t$ does not develop a caustics if $\alpha\geq -\lambda$. The case 
$\alpha=-\lambda$ is special since then $\Lambda_t=V^s$, i.e., $\Lambda_t=\Lambda_0$ is invariant and 
equal to the stable manifold. In this case the manifold is contracting, but for all $\alpha>-\lambda$ 
the manifold is non-contracting. The case $\alpha=\lambda$ is as well special since then 
$\Lambda_t=V^u$ for all $t$, so the initial manifold is the unstable manifold, in  all other cases 
$\Lambda_t\to V^u$ for large $t$. These different  cases are as well reflected in the 
behaviour of $T(t)$ and $ M(t)$. For $\alpha=-\lambda$ we have 
$\phiL(x)=\ue^{-\lambda t}x$ and hence 
\begin{equation}
T(t)A(x)=\ue^{\lambda t/2}A(\ue^{\lambda t}x)\,\quad\text{and}\quad \tilde M(t)=\ue^{\frac{\ui\hbar}{2}\frac{\ue^{2\lambda t}-1}{2\lambda}\Delta}\,\, .
\end{equation}
The transport operator $T(t)$ is squeezing at an exponential rate, and the metaplectic correction 
has to make up for this at an exponential rate, too. In contrast, if $\alpha>-\lambda$ we have 
\begin{equation}
\alpha \lambda^{-1} \sinh(\lambda t)+\cosh(\lambda t)=\frac{\alpha+\lambda}{2\lambda}\ue^{\lambda t}+O(\ue^{-\lambda t})
\end{equation}
and so $T(t)A(x)= [\phiL'(t)]^{-1/2}A(\phiL^{-1}(t,x))$ is stretching at an exponential rate. The metaplectic 
correction is given by 
\begin{equation}
 M(t)=\ue^{\frac{\ui\hbar}{2}[\frac{2\lambda}{(\alpha+\lambda)^2}+O(\ue^{-\lambda t})]\Delta}
\end{equation}
and tends exponentially fast to a limit. This dichotomy between the cases $\alpha=-\lambda$ and $\alpha>-\lambda$ is 
analogous the the one between $\alpha =0$ and $\alpha>0$ we found for the free particle. For $\alpha>-\lambda$ the 
metaplectic correction saturates in time, whereas for $\alpha=-\lambda$ its contribution grows exponentially. 
Since the Hamiltonian is quadratic both cases give exact solutions, and the different initial manifolds 
only correspond to different ways to split the time evolution. But as in the integrable case, if we move to a perturbation 
the contracting case $\alpha=-\lambda$ gives remainder terms which blow up, whereas the expanding 
cases $\alpha>-\lambda$ give remainder terms which remain bounded by $\sqrt\hbar$ independent of time.

So far we have looked at an initial state which is concentrated on top of the barrier, let us now look at a initial state 
localised at $(p,q)$ wit $q<0$ and $p>0$, hence an incoming state. The question of interest is then if this state
gets transmitted or reflected, and how one can describe the transition between reflection and transmission uniformly 
 if one changes $(p,q)$. We chose an initial phase function 
$S_0(x)=p(x-q)+\alpha(x-q)^2/2$, which again for $\alpha >-\lambda$ gives an 
expanding initial manifold $\Lambda_0$. For simplicity we will choose  $\alpha=\lambda$, i.e., $\Lambda_0=(p,q)+V^u$, 
the general case $\alpha>-\lambda$ can be treated similarly. From computing $\Phi^t(\nabla S_0(x),x)$ we find 
\begin{equation}
\phiL(t,x)=\ue^{\lambda t}(x-q)+q(t)\,\, ,
\end{equation}
where $q(t)=\lambda^{-1}p\sinh(\lambda t)+q\cosh(\lambda t)$, and therefore 
\begin{equation}\label{eq:T_par_bar}
T(t)L_qa(x)=\frac{\ue^{-\lambda t/2}}{\hbar^{1/4}}a\bigg(\frac{\ue^{-\lambda t}}{\sqrt{\hbar}}\big[x-q(t)\big]\bigg)\,,\, \quad 
M(t) =\ue^{\frac{\ui}{2}\frac{1-\ue^{-2\lambda t}}{2\lambda} \Delta}\,\, .
\end{equation}
For the phase function we find 
\begin{equation}
S(t,x)=\mathcal{L}(t)+p(t)(x-q(t))+\lambda (x-q(t))^2/2\,\, ,\quad\text{where}\quad (p(t),q(t))=\Phi^t(p,q)\,\, ,
\end{equation}
and $\mathcal{L}(t)=\int_0^t (p\dot q-H(p,q))\, \ud t$. We see that $M(t)$  again tends with exponential speed to a limit 
$M^{\infty}= \ue^{\frac{\ui}{2}\frac{1}{2\lambda} \Delta}$. 
The behaviour of $T(t)L_qa$ depends crucially on $\ue^{-\lambda t}/\sqrt\hbar$, for small time this 
parameter is large, hence the state is localised, but for 
\begin{equation}
\ue^{-\lambda t}\approx \sqrt{\hbar}
\end{equation}
i.e., on Ehrenfest time scales, the function $T(t)L_qa$ is no longer strongly localised and the 
transported wave packet is actually a Lagrangian state associated with 
\begin{equation}
\Lambda_t=\Phi^t(\Lambda_0)=(p(t),q(t))+V^u\,\, .
\end{equation}
Where this state moves depends on $q(t)$, and for long times we have 
\begin{equation}
q(t)=\frac{p+\lambda q}{2\lambda}\ue^{\lambda t}+O(\ue^{-\lambda t})
\end{equation}
so the behavior of $q(t)$ depends on the value of $p+\lambda q$. The case 
$p+\lambda q=0$ corresponds to initial conditions $(p,q)$ in the stable manifold $V^s$ 
 of the fixed point at $(0,0)$, and so the trajectory runs into the fixed point and the 
 Lagrangian state is for long times located on the unstable manifold $V^u$. 
 The case $p+\lambda q>0$ gives a wave packet which is transmitted over the barrier and 
 the case $p+\lambda q <0$ corresponds to a wave packet which is reflected. 
 The strength of our approach is that we can describe the transition between 
 these different cases in a uniform way given by \eqref{eq:T_par_bar}. 
 We studied a simple example here, but the method works for general 
 potential barriers and allows to describe the transition between reflection and 
 transmission of  time dependent wave packets in a uniform way.  
 
 Such processes are of great importance in the theory of chemical reactions and 
 we expect that our method could be of great use in the theoretical and 
 quantitative description of time resolved chemical reactions which are 
 experimentally accessible in femto and atto chemistry \cite{RicZha00,CorKra07}. A first step will be 
 to analyse the transport of wave packets over barriers, or more 
 generally through bottlenecks in phase space, using the method of 
 Quantum Normal Forms  recently developed in \cite{SchuWaaWig06, SchuWaaWig08}.


\section{Transport along a hyperbolic trajectory and its unstable manifold}
\label{sec:exII}

\begin{figure}[htp]

\centering 
\includegraphics[angle=0.0, width=14cm]{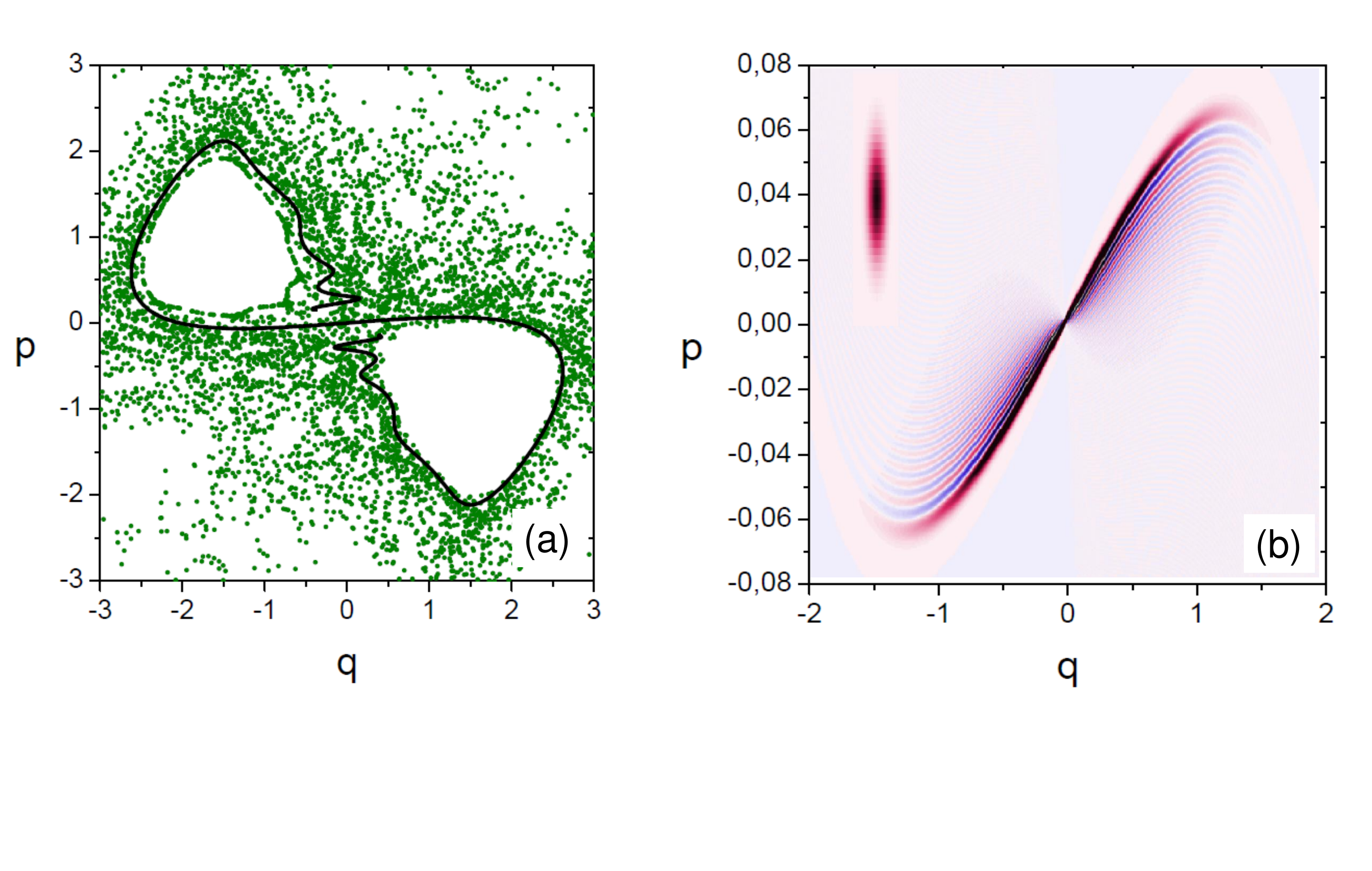}
\vspace{-2cm}
\caption{
In (a) we show a phase space portrait  of the KHO dynamics, \eqref{eq:KHO},  for $K=2$.
The origin is a hyperbolic fixed point, and we displayed as well the  corresponding unstable  manifold.
In (b) we plot the Wigner function of the evolved state  \eqref{eq:init_state} at $t=4^-$, which is of the order 
of the Ehrenfest time $T_E= 4.29\dots  $. It  evolves along the unstable manifold and starts resembling a WKB state. For comparison in the upper left corner the Wignerfunction at $t=0$ is shown, note that  the vertical axis has been rescaled compared to (a).   }
\label{fig1}
\end{figure}

In this section we want to discuss the case that the initial coherent state is concentrated in a point $z=(p,q)$ on a 
hyperbolic trajectory $z(t)=\Phi^t(z)$.  As a test system we choose the  kicked harmonic oscillator 
\begin{equation}\label{eq:KHO}
H(t,p,q,t) =
\frac{1}{2} \left( p^2 +  q^2 \right)
+ K \cos q 
\sum_{n = -\infty}^{\infty} \delta( t - n  )\, ,                   
\end{equation}
where $K$ is the chaoticity parameter which we will choose to be $K=2$. 
This system has as well been used in \cite{MaiNicValTos08}.
In part (a) of Figure~\ref{fig1} we show  a portrait of the classical phase space, the system has a 
an unstable fixed point at the origin and we display the  unstable manifold. The Liapunov exponent of the unstable fixed point is 
$\lambda=0.83...$ and we use $\hbar=0.0008$, so the Ehrenfest time is 
\begin{equation}
T_E=4.29... 
\end{equation}
%
%

\begin{figure}[htp]
\centering 
\includegraphics[angle=0.0, width=14cm]{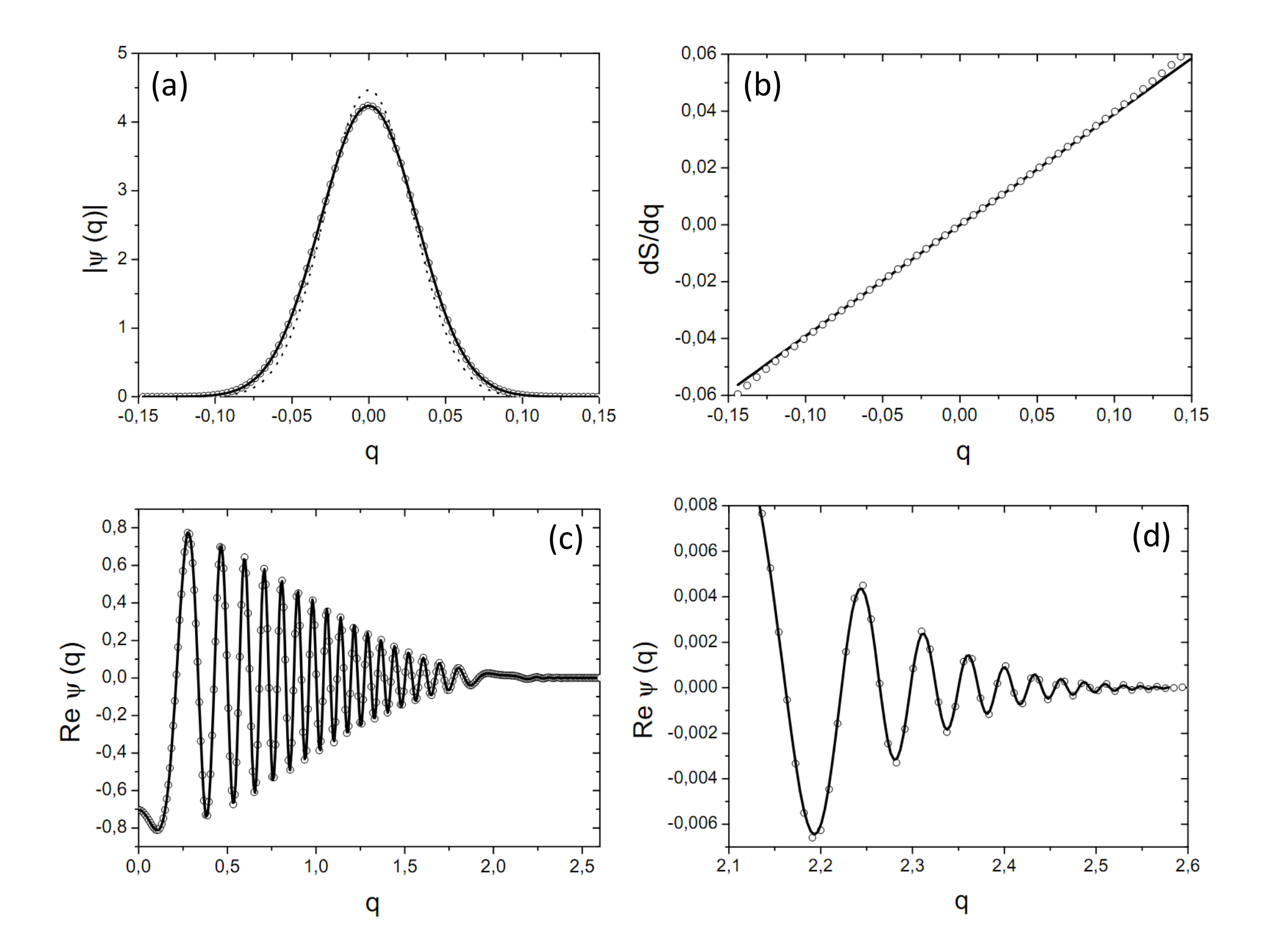}
\caption{%
In the top panel we test \eqref{eq:prop_forw_back}: The initial state \eqref{eq:init_state} is propagated forward 
using the full propagator to $t=4$ and propagated backwards using just time dependent WKB, 
the result (symbol) is compared with the  metaplectic 
approximation (full line): (a) amplitude, (b) phase derivative.
We also show the initial state for reference (dotted lines
in (a)). 
In the bottom panel we compare the metaplectically extended WKB scheme with exact quantum propagation
(circles).  The right panel (d) is a zoom of the tail of the wavefunction
in (c) to show that agreement extends down to this scale and
into the most nonlinear region }
\label{fig2}
\end{figure}

As initial state we choose  a Gaussian wavepacket centred on the 
fixed point at the origin:
\begin{equation}\label{eq:init_state}
\psi_0(x)= \frac{1}{ (\pi\hbar)^{1/4} } \, e^{ -x^2/2 \hbar} \,  .           
\end{equation}
In Fig.~\ref{fig1} (b) we display the Wigner function of the time evolved state at $t=4$ which 
can be seen to be stretched along the unstable manifold. As expected at the Ehrenfest time 
the state becomes a WKB type state associated with the unstable manifold. 
For comparison  the real part of the time evolved 
wave function $\psi(t)$ is plotted in  part (c) of Fig.~\ref{fig2}.  

%
%

The prediction of the theory in 
Section \ref{sec:met_ext}
is that, after a suitable
metaplectic correction of the initial amplitude, the state can
be propagated using standard time dependent WKB approximation .
This means that, if we decompose the initial state as
\begin{equation}
\psi_0 (x) = (L_0a)(x) \, \ue^{i S_0 (x)/\hbar}  \, , 
\label{eq:fiducial}          
\end{equation}
then the evolved state is by \eqref{eq:TDS} and \eqref{eq:TM_approx} given by
\begin{equation}
\psi(t,x) 
=\left[ T(t) \, D(t) \, L_0a_0 \right](x)  e^{i S(t,x)/\hbar} 
 \approx 
\left[ T(t) \, M_0(t) \, L_0a \right] (x) e^{i S(t,x)/\hbar}  \, .
\label{eq:wkbplus}
\end{equation}
Here $S(t,x)$ is the solution of the Hamilton-Jacobi equation
satisfying $S(0,x)=S_0(x)$.
The operators $T(t)$, $D(t)$ and $M_0(t)$ also depend on the 
choice of $S_0(q)$. 
Even if according to Section \ref{sec:initial_man} this choice is, to some extent, arbitrary, the efficiency
of the method may be sensitive to $S_0(x)$.
We shall only consider the quadratic case  $S_0(x)=\frac{\tan\theta}{2}\, x^2$, 
corresponding to a linear Lagrangian manifold
$p=\tan(\theta)\,  q$.
We shall see that there is a wide range of 
initial manifolds that work well.

Note that Eq.~(\ref{eq:wkbplus}) is equivalent to saying that 
the sequence of operations
\begin{equation}\label{eq:prop_forw_back}
 T^\ast(t) \left[ \psi(t) \, e^{-i S(t)/\hbar} \right] 
 =D(t)L_0\, a_0\approx M(t)L_0 \, a_0 \,
\end{equation}
must produce approximately a Gaussian state centred at $0$, 
for a Gaussian initial amplitude $a_0(x)$.
This is the first test we shall carry out, with the choice $S_0=0$  corresponding to
the initial Lagrangian manifold $p=0$.
First we propagate the state (\ref{eq:fiducial}) 
exactly during some time, and then propagate it back using time dependent WKB.
Panel (a) and (b) in Figure~\ref{fig2} shows the comparison of the exact
result $D(t)\,L_0a_0$ with the metaplectic approximation
$M_0(t)\,L_0a_0$.
In the case of the modulus the agreement between the exact
$DL_0a_0$ and $M_0L_0a_0$ is perfect within visual resolution.
There is a small difference between the phases, but this 
occurs in a region where the amplitude is very small. 
We  show as well a direct  comparison between
the exact and the metaplectically extended WKB propagated states in 
part (c) and (d) of  Fig.~\ref{fig2}. 
The agreement is excellent.

\begin{figure}[htp]
\centering 
\includegraphics[angle=0.0, width=14cm]{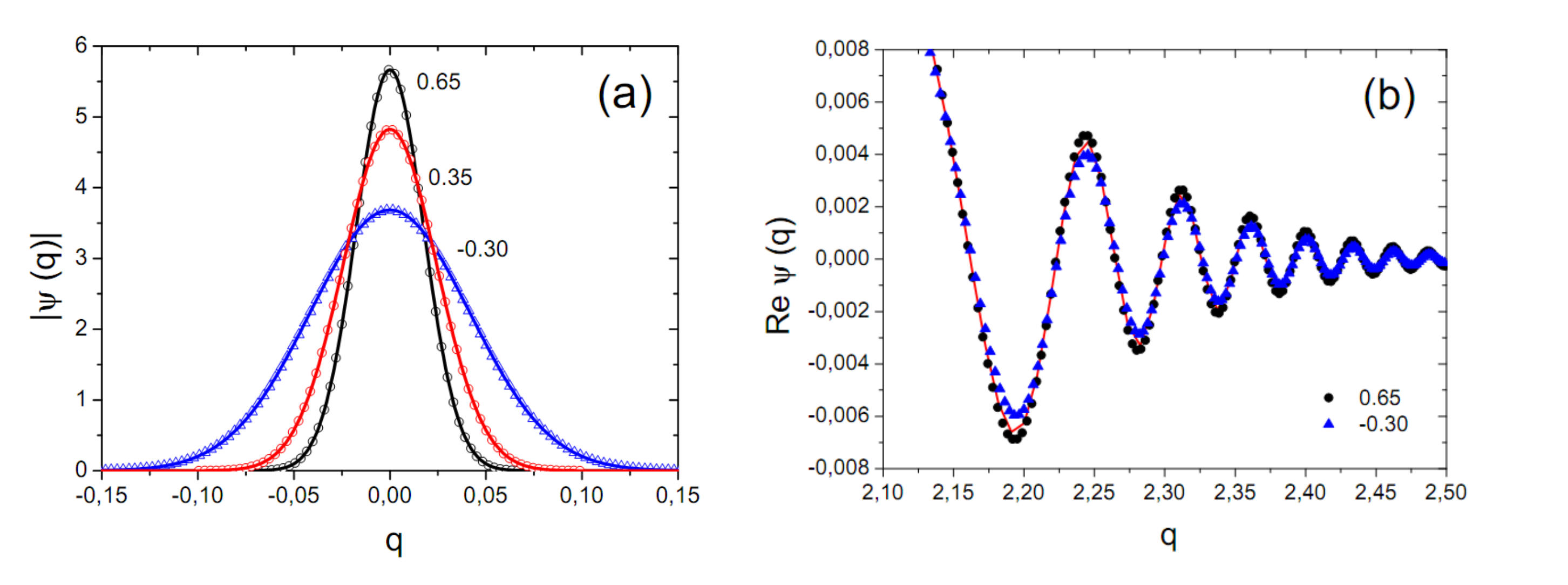}
\caption{%
We compare the effect of different initial Lagrangian manifolds $p=\tan\theta\, q$ on the propagation, we considered 
for $\theta$ the following fractions of $\pi/2$: -0.30, 0.35, 0.65.
 On the left 
we display  $\abs{DL_0a_0}(q)$ (circles) and  the metaplectic approximations $\abs{M_0L_0 a_0}(q)$ (lines). 
On the right we show the evolved state at $t=4^-$ (real part).
We compare metaplectically extended  WKB (symbols: black dots and blue triangles) with exact quantum propagation
(line), two slopes where considered: $0.65 \pi/2$ and $-0.30 \pi/2$}
\label{fig3}
\end{figure}

We will now investigate how  robust the method is with respect
to the changing the choice of the (linear) initial WKB manifold.
We have changed the slope $\theta$ in a range of almost 90 degrees,
from $\theta =-0.30 \pi/2$ to $\theta=0.65 \pi/2$.
The comparison between exact and metaplectically extended WKB 
 at $t=4$ are shown in Fig.~\ref{fig3}.
Except for very slight differences (blue triangles) the
agreement is still excellent for these ``large" slopes.

%
%

As we increase the slope, the amplitude concentrates in 
narrower regions. 
This happens because, the larger the slope, 
the map induced in the $q$-coordinate by the WKB manifold is 
more expansive, and $T^\ast$ is more contracting.

We tested the theory as well for different values for $\hbar$ and found good agreement 
with the expected behaviour (not shown).


\section{Conclusions}

In this paper we derived an extension of the standard time dependent WKB method which can be applied to 
highly localised states like coherent states. It allows to describe in a uniform way the transition in time from a semiclassically 
highly localised coherent state to an delocalised  Lagrangian state which takes place at the Ehrenfest time.

The main idea on which this extension of time dependent WKB theory is built is an exact decomposition of the time 
evolution of an initial state of the form $\psi_0(x)=A_0(x)\ue^{\frac{\ui}{\hbar}S_0(x)}$, where $S_0$ is real valued, 
into several parts
\begin{equation}
\psi(t)=(T(t)D(t)A_0)\ue^{\frac{\ui}{\hbar}S(t)}\,\, .
\end{equation}
Here $S(t)$ is a solution of the Hamilton Jacobi equation and is hence related to the 
transport of the Lagrangian manifold $\Lambda_0$ generated by $S_0$ through phase space. 
The unitary operator $T(t)$ transports functions in position space along the 
projections of the phase space trajectories emanating from $\Lambda_0$.  Finally 
$D(t)$ is the propagator generated by the time dependent Hamiltonian $-\hbar^2 T^*(t)\Delta T(t)/2$. 
$S(t)$ and $T(t)$ are defined purely in terms of transport along classical trajectories and 
$D(t)$ takes into account the dispersive effect of quantum mechanics. 

The standard time dependent WKB approximation is obtained by approximating $D(t)\approx I$, this works fine if 
the amplitude $A_0$ is sufficient flat, i.e., has bounded derivatives. 
But for a coherent state $A_0$ is strongly localised around a point $x=q$, then 
it is more natural to freeze the coefficients of  the generator of $D(t)$ at $x=q$, the resulting operator is a metaplectic operator 
$M_q(t)$ whose action on functions can be computed quite easily. The main result of this paper is thus  
\begin{equation}
\psi(t)\approx (T(t)M_q(t)A_0)\ue^{\frac{\ui}{\hbar}S(t)}\,\, .
\end{equation}
which is valid for amplitudes $A_0$ which are strongly localised around $x=q$.  In order to justify the validity of this 
approximation for long times, particularly on Ehrenfest time scales, we had to 
analyse the underlying classical dynamics more carefully. We introduced a non-contraction condition on the 
position space trajectories  emanating from a neighbourhood of the initial state with momenta $p=\nabla S(x)$, and 
if this condition holds our propagation scheme is effective. The non-contraction condition excludes caustics, but 
we have a large freedom in the choice of the initial phase function $S(x)$ which allows in many cases to avoid caustics, 
at least until the  state becomes delocalised. 

For times shorter than the Ehrenfest time our scheme reproduces the standard coherent state propagation 
results based on a Taylor expansion of the Hamiltonian around the centre trajectory. But for times 
of the order of the Ehrenfest time we find that the state becomes extended and a Lagrangian state, 
which in the chaotic case is supported by the unstable manifold of the centre trajectory. From that time onwards standard 
time dependent WKB theory applies, as has been observed in \cite{MaiNicValTos08}. In particular if the system is 
hyperbolic and mixing one can apply the results from \cite{Schub05}  to conclude the state becomes equidistributed after the 
Ehrenfest time.

In order to extend the results to more general Hamiltonians, and to be able to include caustics, we noticed that 
 standard time dependent WKB approximation can be viewed as the exact quantum time evolution generated by a 
 quantisation of a first order Taylor approximation of the Hamilton function around the Lagrangian manifold 
 associated with the time evolved  WKB state. Based on this insight we could now choose different 
 first order approximations which remained valid for general Hamiltonians and at caustics. The price one has  
  to pay is a more complicated and less explicit formalism. In addition all the 
  previous results were in principle rigorous, although we refrained from stating them in the form of theorems, but 
  here we have to rely on an assumption that certain special cases of Egorov's theorem remain valid 
  on Ehrenfest time scales. But we get a further benefit from this more general way 
  to look at the time dependent WKB approximation, we can show that the classical map associated with 
  the  metaplectic correction $M_q(t)$ is a shear map on phase space and that it furthermore in many cases  
  converges to a limit for large $t$ which can be determined from simple geometric considerations. This implies that
   in many cases 
   \begin{equation}\label{eq:Mto}
   \lim_{t\to\infty} M_q(t)=M_q^{(\infty)}
   \end{equation}
   and we can determine $M_q^{(\infty)}$ up to a phase from simple geometric considerations. If the centre trajectory is 
   hyperbolic then the limit in \eqref{eq:Mto} is reached exponentially fast.    
  
  We illustrated and tested the theory with several examples. For the free particle, and more generally, for integrable systems, 
  the dynamics can be computed quite explicitly. The Ehrenfest time is of order $T_E\sim \hbar^{-1/2}$ and we see 
  a qualitative transition in the nature of the state from a localised coherent state  to an extended  
  WKB state at this time scale. Similarly for a parabolic barrier the dynamics can be 
  computed explicitly and the formalism developed in this works gives explicit formulas which describe in a uniform way the 
  transition from reflection to transmission of a wavepacket when one varies the energy near the critical energy. 
  We then carried out detailed numerical tests on the kicked harmonic oscillator for an initial state localised on 
  a hyperbolic fixed point.  These showed impressive agreement  between the metaplectic extension of the 
  WKB method and exact quantum propagation. These tests illustrated as well the simplicity of the method;
  the metaplectic correction is in fact a very simple operator, and if one has implemented the 
  time dependent WKB method then adding the metaplectic correction is easy and allows at once  to 
  propagate a much larger class of states.

  There are  many areas where  the results from this paper should be of interest. In many physical system the 
  Ehrenfest time is for a realistic set of parameters actually quite short. E.g., in quantum   billiards 
  coherent states spread out after a few bounces, \cite{TomHel91,TomHel93}, and the metaplectic extension of 
  WKB should be able to describe this transition. The scattering of a wavepacket of a barrier near the critical energy is 
   another example, the Ehrenfest time is the time when the wavepacket reaches the barrier, i.e., the time where the 
   physically interesting processes start to happen. Metaplectically extended WKB allows to describe this process 
   explicitly and uniformly  in $\hbar $ and $t$.  A large class of chemical reactions is described in the 
   framework of Transition State Theory by the crossing of a barrier on a high dimensional energy surface and 
   modern experimental methods in atto and femto chemistry allow to study the dynamics of such chemical 
   reactions with impressive precision, \cite{RicZha00,CorKra07}. The metaplectic extension to WKB together with the normal form approach to 
   Transition state theory, \cite{SchuWaaWig06, SchuWaaWig08}, should allow to give an efficient theoretical description of 
   chemical processes on such time scales.

\appendix

\section{Wigner Weyl correspondence}

In this appendix we collect some material on Weyl quantisation
which we use in the main part. Most of it is a development 
of standard material and only the result on Egorov's theorem 
seems to be new. References for the material we present are \cite{DimSjo99,Dui11}.

\subsection{Weyl quantisation} 

Let $\hat W(\xi, x):=\ue^{\frac{\ui}{\hbar} (\xi \hat q+x\hat p)}$ be the 
Weyl operators which represent translations in phase space, then we 
can define for a function $A(p,q)$ on phase space an operator, its \emph{Weyl quantisation}, by 
\begin{equation}
\hat A=\iint \mathcal{F}A(\xi,x)\hat W(\xi,x)\frac{\ud x\ud \xi}{(2\pi\hbar)^{2n}}
\end{equation}
where $\mathcal{F}A(\xi,x)=\iint A(p,q)\ue^{-\frac{\ui}{\hbar}[p\xi+qx]}\, \ud p\ud q$ denotes the Fourier 
transform of $A$.  The function $A(p,q)$ is  called the \emph{Weyl symbol} of $\hat A$, 
and the properties of $\hat A$ can often be determined from properties of $A$, e.g., if $A\in S'(\R^n\times \R^n)$, 
then $\hat A :S(\R^n)\to S'(\R^n)$. If $\hat A=|\psi\ra\la \psi |$ is the projection onto 
a state $\psi$, then $A(\xi,x)$ is proportional to the Wigner function of the state 
$\psi$.  

The product of operators can be expressed in terms of the symbols, one has 
$\hat A\hat B=\widehat{A\sharp B}$ where 
\begin{equation} \label{eq:symb-prod}
\begin{split}
A\sharp B(p,q)=&A(p,q)\ue^{\frac{\ui \hbar}{2}[\overleftarrow\nabla_p\cdot\overrightarrow\nabla_q-\overleftarrow\nabla_q\cdot \overrightarrow\nabla_p]}B(p,q)\\
=&A(p,q)B(p,q)+\frac{\ui \hbar}{2} \{A,B\}(p,q)\\
&-\frac{\hbar^2}{8}A(p,q)[\overleftarrow\nabla_p\cdot\overrightarrow\nabla_q-\overleftarrow\nabla_q\cdot \overrightarrow\nabla_p]^2B(p,q)+\cdots\,\, .
\end{split}
\end{equation}
Here $\{A,B\}(p,q)$ denote the Poisson bracket, and the arrows over the derivatives indicate if they act on the function
on the left or on the right. Suitable conditions on the functions $A$ and $B$ under which this expansion holds can  be found 
in \cite{DimSjo99}. 

\subsection{Dynamics generated by first order operators}
\label{app:vf}

Let $K(t,x,\xi)=X(t,x)\cdot \xi$,  where $X(t,x)$ is a time dependent 
vector field, we are interested in the time evolution $T(t,s)$ generated by 
\begin{equation}
\hat K(t)=-\ui\hbar X(t,x)\cdot \nabla -\frac{\ui\hbar}{2}\nabla X(t,x)\,\, .
\end{equation}
Let $\phi(t,s,x)$ be the family of maps generated by the 
vector field $X(t,x)$, i.e., 
\begin{equation}
\pa_t\phi(t,s;x)=X(t,\phi(t,s;x))\,\, ,\quad \phi(t,t;x)=I\,\, , 
\end{equation}
then 
\begin{equation}
(T(t,s)A)(x)=[\det \phi'(t,s;x)]^{-1/2}A(\phi^{-1}(t,s;x))\,\, .
\end{equation}

\subsection{Egorov's theorem}
\label{app:egorov}

Egorov's theorem is one way to formulate the correspondence principle, 
it gives a general relation between classical and quantum dynamics. 
Let $H(t)$ be a real valued smooth phase space function,  $\hat H(t)$ its 
Weyl quantisation and $\Phi^t$ and  $U(t)$ be the classical and quantum time evolution generated by 
$H(t)$ and $\hat H(t)$, respectively.  Then Egorov's theorem states  that 
\begin{equation}
U^*(t)\hat AU(t)=\hat A_t +O_t(\hbar^2)\,\, ,
\end{equation}
where $A_t=A\circ \Phi^t$,  and $A$ and $H$ have to satisfy some conditions on their smoothness and growth at infinity, 
see \cite{BouRob02}. Since $A_t$ is the classical time evolution of $A$ this means that quantum and 
classical evolution are close for small $\hbar$. The remainder term $O_t(\hbar^2)$ does depend on time, 
and the best general estimates are of the form $O_t(\hbar^2)\ll \hbar^2 \ue^{\Gamma t}$ for some constant  $\Gamma>0$ 
which depends on $H$. We want to discuss two cases in which one has better control over the remainder. 

One approach to Egorov's theorem is based on writing Heisenberg's equation of motion for 
$\hat A(t)=U^*(t)\hat AU(t)$, i.e., $\ui\hbar \pa_t\hat A(t)=[\hat H,\hat A]$, in terms of the symbols using 
\eqref{eq:symb-prod} which gives 
\begin{equation}\label{eq:WigHei}
\begin{split}
\pa_t A(p,q)&=\frac{2}{\hbar}A(p,q) \sin(\hbar [\overleftarrow\nabla_p\cdot\overrightarrow\nabla_q-\overleftarrow\nabla_q\cdot \overrightarrow\nabla_p]/2)H(p,q)\\
&=\{A,H\}(p,q)+\frac{\hbar^2}{24}A(p,q)[\overleftarrow\nabla_p\cdot\overrightarrow\nabla_q-\overleftarrow\nabla_q\cdot \overrightarrow\nabla_p]^3 H(p,q)+\cdots \,\, .
\end{split}
\end{equation}
The leading order term is just the Liouville equation and gives $A\approx A\circ\Phi^t$, the main problem is then to control
the higher order terms. 

\subsubsection{Quadratic Hamiltonians and Metaplectic operators}

If $H(p,q)$ is a quadratic function of $p$ and $q$, with possibly time dependent coefficients, 
then the $\hbar$ expansion in \eqref{eq:WigHei} terminates after the leading term, and hence 
the evolution equation for $A$ is just the classical Liouville equation. So in this case 
\begin{equation}
U^*(t)\hat AU(t)=\hat A_t\,\, ,
\end{equation}
where $A_t=A\circ \Phi^t$, and one says Egorov's theorem is exact. 
 The corresponding classical maps $\Phi^t$ are linear, hence 
the operators $U(t)$ for all quadratic $H$ form a quantisation of the symplectic group, which 
turns out to be a double cover of the symplectic group called the metaplectic group. The operators 
$U(t)$ are often referred to as metaplectic operators \cite{Lit86,Fol89,ComRob06}.

\subsubsection{Conjugation by a flow}
\label{app:Eg}
The second case we need is that $U(t)=T(t,0)$, i.e., $H(t)$ is linear in $p$ and given 
by $H(t,p,q)=K(t,p,q)=X(t,q)\cdot p$.  The classical 
map $\Phi_1^t(p,q)$ generated by $H$ is given by the solutions to 
\begin{equation}
\dot \xi(t) =-\nabla_q K(t,\xi(t),x(t))\,\, ,\quad \dot x (t)=\nabla_p 
K(t,\xi(t),x(t)) 
\end{equation}
with initial conditions $\xi(0)=p$ and $x(0)=q$. We can express 
$\Phi^t_1$ in terms of the map $\phi(t,0,x)$ as 
\begin{equation}
\Phi^t_1(\xi,x)=\big([\phi'(0,t,x)]^{\dagger}\xi,\phi(t,0,x)\big)
\end{equation}
where $[\phi'(0,t,x)]^{\dagger}$ is the transpose of the inverse of the matrix
$\phi'(t,0,x)$.

Let us consider first the case that $A$ is linear in $p$, i.e., $A=b(q)\cdot p$ for some 
vector valued function $b(q)$, then only the leading order term in \eqref{eq:WigHei} 
is non-zero, and hence 
\begin{equation}\label{eq:lin_Egorov}
T^*(t,0)\hat  A T(t,0)=\hat A_t\,\, ,\quad\text{with}\quad A(t,q,p)=b(\phi(t,0;q))\cdot [\phi'(0,t,x)]^{\dagger}p\,\, ,
\end{equation}
without any remainder terms, hence Egorov is exact again. Using this result we can discuss the case we 
will need, namely the case that $\hat A$ is a second order differential operator of the  form 
\begin{equation}
\hat A=(\cB(q)\nabla) \cdot \cB(q)\nabla 
\end{equation}
So that $T^*\hat AT=T^*\cB(q)\nabla T\cdot T^* \cB(q)\nabla T$ and if we denote the rows of $\cB$ by $b_j(q)$ we can use the 
previous result \eqref{eq:lin_Egorov} and \eqref{eq:symb-prod} to obtain $T^*(t,0)\hat A T(t,0)=\hat A(t)$ with 
\begin{equation}
\begin{split}
A(t)&=\sum_{i}\big(b_i(\phi(t,0;q))\cdot [\phi'(0,t,x)]^{\dagger}p\big) \sharp \big(b_i(\phi(t,0;q))\cdot [\phi'(0,t,x)]^{\dagger}p\big)\\
&=p \cdot \phi'(0,t,x)(\cB^{\dagger}\cB)(\phi(t,0;q))[\phi'(0,t,x)]^{\dagger}p\\
&\quad+\frac{\hbar^2}{8}\sum_{i} \big(b_i\cdot [\phi'(0,t,q)]^{\dagger}p\big)(\overleftarrow\nabla_p\cdot\overrightarrow\nabla_q)(\overleftarrow\nabla_q\cdot \overrightarrow\nabla_p) \big(b_i(\phi(t,0;q))\cdot [\phi'(0,t,x)]^{\dagger}p\big)
\end{split}
\end{equation}
where  we have used \eqref{eq:symb-prod} . This can be further simplified, but we will restrict  ourselves to the case 
$\cB=I$, hence $b_i=e_i$ and then we find 
\begin{equation}
A(t,p,q)= p\cdot \phi'(0,t,x)[\phi'(0,t,x)]^{\dagger}p
+\frac{\hbar^2}{8}\sum_i \Tr [\phi_i''(0,t,q)]^2\,\, ,
\end{equation}
which is the symbol of $T^*(t,0)\Delta T(t,0)$. The term of order $\hbar^2$ contains second derivatives of 
the inverse of $\phi(t,0;q)$ and so if $\phi(t,0;q)$ is non-contracting, then these derivatives stay bounded. 

\section{A Lemma on Symplectic Maps}
\label{app:symp_lemma}

The results in this Appendix are used at the end of Section \ref{sec:gen_scheme} to show that 
the classical map associated with the metaplectic correction is uniquely determined by the dynamics of 
the tangent space to the initial Lagrangian submanifold and the vertical subspace.

\begin{lem}
Let $L_1,L_2\subset \R^n\times \R^n$ be  Lagrangian subspaces with $L_1\cap L_2=\{0\}$ and let 
$L$ be another Lagrangian subspace with $L\cap L_1=\{0\}$. Then there exist a unique   linear symplectic map $T$ with 
$T|_{L_1}=I$  and $T(L_2)=L$. 
\end{lem}

To show this we will use mainly two standard facts from linear symplectic geometry, see e.g. \cite{Dui11}: 
\begin{itemize}
\item[(a)] By Darboux' Theorem there exist  symplectic coordinates 
$(v,w)\in \R^n\times\R^n$ such that $L_1=\{w=0\}$  and $L_2=\{v=0\}$.
\item[(b)] If $L_3$ is a Lagrangian subspace  with $L_3\cap L_1=\{0\}$ then there exists a unique 
symmetric matrix $A$ such that $L_3=\{v=Aw\}$. 
\end{itemize}

Let us now prove the lemma. The condition $T|_{L_1}=I$ implies that in the coordinates from $(a)$ the 
 matrix representing $T$ is of the form $M_T=\begin{pmatrix} I & A\\ 0 &B\end{pmatrix}$ where 
$A,B$ are $n\times n$ matrices. Now this matrix must be as well symplectic, i.e., $M_T^t\Omega M_T=\Omega$ with 
$\Omega=\begin{pmatrix} 0 & -I\\ I & 0\end{pmatrix} $ and this  gives $B=I$ and $A=A^t$. 
Therefore  
\begin{equation}
T(L_2)=\{v=Aw\}\,\, ,
\end{equation}
but by $(b)$ the condition $T(L_2)=L$ determines then $A$ uniquely.  So $M_T=\begin{pmatrix} I & A\\ 0 &I\end{pmatrix}$ 
is the unique  shear  relative to $L_1$ which maps $L_2$ to $L$.

\section*{References}


\end{document}